\documentclass[notitlepage]{article}

\usepackage[a4paper, total={7in, 9in}]{geometry}
\usepackage{tikz}
\usepackage{caption}
\usepackage{graphicx}
\usepackage{subcaption}
\usepackage[normalem]{ulem}

\usepackage{longtable}
\usepackage{comment}

\usepackage[
    backend=bibtex,
    natbib = true,
    style = phys,
    isbn = false,
    hyperref = true,
    sorting = none,
    giveninits = true,
    maxbibnames = 9,
    biblabel = brackets,
    eprint = true
]{biblatex}

\bibliography{references}

\usepackage{amsmath}
\usepackage{amsthm}
\usepackage{braket}

\usepackage{tikz}
\usepackage{pgfplots}
\graphicspath{{figures/}{.}}
\makeatletter
\def\input@path{{figures/}{.}}
\makeatother

\pgfplotsset{compat=1.16}
\providecommand\inputpgf[2]{{
\let\pgfimageWithoutPath\pgfimage
\renewcommand{\pgfimage}[2][]{\pgfimageWithoutPath[##1]{#1/##2}}
\input{#1/#2}
}}

\usepackage{listings}
\usepackage[ruled,vlined]{algorithm2e}
\usepackage{amssymb}
\usepackage[ampersand]{easylist}
\usepackage{enumitem}
\usepackage{hyperref}
\hypersetup{
  colorlinks = true, 
  linkcolor={blue!50!black}, 
  citecolor=red, 
  urlcolor=blue, 
  linktoc=page,
}
\usepackage[capitalise, nameinlink]{cleveref}
\usepackage[
  style=long,
  nolist, 
  nonumberlist,
  nopostdot,
  acronym,
  toc=true, 
]{glossaries}
\usepackage{bm}
\usepackage{booktabs}
\usepackage{multirow}

\usepackage{array}
\usepackage[british]{babel}

\newtheorem{theorem}{Theorem}
\newtheorem{proposition}[theorem]{Proposition}
\newtheorem{lemma}[theorem]{Lemma}
\newtheorem{corollary}[theorem]{Corollary}

\usepackage{makecell}

\usepackage{xspace}

\let\originalleft\left
\let\originalright\right
\renewcommand{\left}{\mathopen{}\mathclose\bgroup\originalleft}
\renewcommand{\right}{\aftergroup\egroup\originalright}

\newcommand{\be}{\begin{equation}}
\newcommand{\ee}{\end{equation}}

\newcommand{\THR}{THRIFT\xspace}
\newcommand{\apx}{apx\xspace}
\newcommand{\FH}{Fermi-Hubbard\xspace}
\newcommand{\TFIM}{transverse-field Ising model\xspace}
\newcommand{\HTFIM}{H_{\textnormal{TFIM}}}
\newcommand{\HFH}{H_{\textnormal{FH}}}
\newcommand{\HHeisenberg}{H_{\textnormal{Heisenberg}}}
\newcommand{\T}{\mathcal{T}}
\newcommand{\hop}{t_{\mathrm{hop}}}

\DeclareMathOperator{\ad}{ad}
\DeclareMathOperator{\diam}{diam}

\newcommand\norm[1]{\lVert#1\rVert}

\usepackage{authblk}
\title{Efficient and practical Hamiltonian simulation \\ from time-dependent product formulas}
\date{\today}

\author[1,2]{Jan Lukas Bosse}
\author[1,3]{Andrew M. Childs}
\author[1]{Charles Derby}
\author[1]{Filippo Maria Gambetta}
\author[1,2]{Ashley Montanaro}
\author[1]{Raul A. Santos}

\affil[1]{Phasecraft Ltd.}
\affil[2]{School of Mathematics, University of Bristol}
\affil[3]{Department of Computer Science, Institute for Advanced Computer Studies, and Joint Center for Quantum Information and Computer Science, University of Maryland}
 
\begin{document}

\maketitle

\begin{abstract}
In this work we propose an approach for implementing time-evolution of a quantum system using product formulas. The quantum algorithms we develop have provably better scaling (in terms of gate complexity and circuit depth) than a naive application of well-known Trotter formulas, for systems where the evolution is determined by a Hamiltonian with different energy scales (i.e., one part is ``large'' and another part is ``small'').
Our algorithms generate a decomposition of the evolution operator into a product of simple unitaries that are directly implementable on a quantum computer. 
Although the theoretical scaling is suboptimal compared with state-of-the-art algorithms (e.g., quantum signal processing), the performance of the algorithms we propose is highly competitive in practice. 
We illustrate this via extensive numerical simulations for several models. For instance, in the strong-field regime of the 1D \TFIM, our algorithms achieve an improvement of one order of magnitude in both the system size and evolution time that can be simulated with a fixed budget of 1000 arbitrary 2-qubit gates, compared with standard Trotter formulas.
\end{abstract}

\section{Introduction}

Time-dynamics simulation (TDS) of quantum systems has long been considered as a natural application where quantum computers can outperform classical ones. 
A quantum algorithm for TDS approximates the time-evolution operator $e^{-itH}$ by a sequence of elementary gates.
The gate complexity of this decomposition is at least linear in $t$ in general \cite{Berry_2006,Childs2010}, and several methods have been proposed that achieve (or nearly achieve) that complexity \cite{Childs_2009,LCU2012,Berry_2014,Low2017}.
These methods differ in the way they implement time evolution, have different overheads, and scale differently with the desired accuracy. 

Arguably the most straightforward TDS algorithm is the use of (Trotter) product formulas~\cite{Ostmeyer_2023_trotter}.
This approach does not use ancilla qubits~\cite{Childs_2009,LCU2012,Berry_2014}, nor does it involve potentially costly operations (such as block encodings or reflections about ancillary quantum states)~\cite{Low2017}, or any classical pre-processing (such as searching for classically optimized circuits~\cite{McKeever_2023_optimized, Tepaske_2023_optimal}).
Moreover, product formulas can be more efficient in practice when simulating systems with hundreds of qubits for times that scale with the size of the system \cite{Childs2018}. This may be due to 
overheads that some asymptotically better algorithms incur, and to the fact that product formula methods scale better in practice than naive bounds suggest, with dependence on commutators of terms that can naturally take advantage of spatial locality \cite{Childs_2019,Childs2021}. 

Product formulas split the evolution under a Hamiltonian $H = \sum_k h_k$ into a product of the form $\prod_{jk}e^{-it_{jk}h_k}$ 
for some times $t_{jk}$. This provides an efficient simulation if each elementary exponential can be implemented efficiently. Observe that the choice of the summands that compose $H$ is not unique. A common practice when simulating lattice systems is to represent the Hamiltonian as a sum of Pauli terms $H=\sum_k\alpha_kP_k$ and choose $h_k=\alpha_kP_k$.  

In this work, we introduce several algorithms that take advantage of the structure of the Hamiltonian to achieve better error scaling than standard product formulas. This approach can leverage knowledge of the gates that can be efficiently implemented in practice on a particular quantum computer, so we call this family of algorithms Trotter Heuristic Resource Improved Formulas for Time-dynamics (\THR).

Starting from a Hamiltonian generating time evolution that can be implemented with a quantum circuit with error independent of the evolution time
(e.g., a Hamiltonian diagonal in the computational basis, or diagonalisable with a circuit that does not scale with the evolution time), we ask, ``What is the effect of adding a perturbation to the Hamiltonian in the complexity of implementing the TDS algorithm with product formulas?". This motivates going into the interaction picture and approximating the time-ordered operator by a product of exponentials. Reference~\cite{low2019hamiltonian} explored such an interaction-picture approach, studying approximations of the time-ordered operator through a Taylor expansion of the Dyson series, instead of using product formulas. We comment on this difference below, when we compare \THR with previous approaches.

\THR generates an efficient product-formula decomposition for time evolution of a quantum system. This decomposition has provably better scaling of both gate complexity and circuit depth than a naive application of well-known product formulas, for systems where the evolution is determined by a Hamiltonian with different energy scales (i.e., in which one part is ``large'' and another part is ``small'', with the size of the small part quantified by a parameter $\alpha$). This situation is ubiquitous in effective models describing physical systems and can occur, for example, in systems with strong short-range interactions and weaker long-range interactions. Furthermore, weak (or strong) external perturbations can be used to push a system out of equilibrium and to extract its dynamical properties.
Crucially, the efficiency of the algorithm depends on the characteristics of the quantum computer itself, namely, the set of gates that are easily implementable with an error independent of the circuit depth.
This is particularly useful in a Noisy Intermediate-Scale Quantum (NISQ) computer, where some types of gates can be implemented more easily than other nominally similar gates.
As these formulas provide better gate complexity than naive product formulas in many instances, we expect them to be useful beyond NISQ applications as well.

In \cref{sec:motivation} we introduce \THR and show that its error scales as $O(\alpha^2t^2)$, an improvement by a factor of $\alpha$ compared with standard
first-order product formulas. We show that $k$th-order \THR achieves error-scaling of $O(\alpha^2t^{k+1})$, compared to $O(\alpha t^{k+1})$ for standard $k$th-order formulas. 
We also show (in \cref{appendix_2}) that general product formulas based directly on products of the summands of the Hamiltonian cannot achieve better scaling than $\alpha^2$. To improve the 
$\alpha$-scaling for higher-order formulas, in \cref{sec:magnus,sec:fer} we introduce the Magnus-\THR and Fer-\THR algorithms, respectively, which achieve an effective $O(\alpha^{k+1} t^{k+1})$ error scaling, for any $k\in \mathbb{N}$.

To complement our theoretical results that show favourable asymptotic scaling of the algorithms, in \cref{sec:numerical_results} we carry out numerical experiments comparing several product formulas with \THR. We analyse the error as a function of the total evolution time and the scale of the small part of the Hamiltonian $\alpha$ for three different models: the transverse-field Ising model in one (1D) and two dimensions (2D), the 1D Heisenberg model with random fields, and the 1D Fermi-Hubbard model. For the spin models studied, the \THR approach generates better product formulas in terms of gate complexity (measured as the number of CNOT or arbitrary 2-qubit gates to achieve a target error) for a wide range of evolution times and $\alpha$. In the case of the transverse-field Ising and Heisenberg models, surprisingly, the complexity is better even when the interaction is stronger than the transverse field. Despite the simplicity of such models, they correctly describe the relevant physics of a wide range of materials and, in the presence of frustration, they can host exotic quantum phases of matter~\cite{Manousakis_1991_heisenberg, Fazekas_1999_magnetism,Zhou_2017_spinliquids}. In these cases, the favourable scaling is due to the possibility of implementing the elementary evolution gates with a 2-qubit gate cost that is the same as standard product formulas. 
For simulations of the Fermi-Hubbard model, \THR methods have advantageous scaling for large enough simulation time 
$T \gtrsim U^{-1}$ and small scale of the hopping term $\hop/U$.
This is due to the extra cost incurred in the implementation of \THR in this case.

\subsection*{Comparison with previous approaches}

Reference \cite{low2019hamiltonian} considers the time evolution of systems with different energy scales and proposes carrying out the simulation in the interaction picture through a method called \emph{linear combination of unitaries} (LCU), achieving a gate complexity of $O(\alpha T\ {\rm polylog}(T\alpha/\epsilon)) $ for a simulation for time $T$ with error $\epsilon$.
Although, theoretically, the LCU method has better scaling with evolution time and simulation error than product formulas, it has also been shown empirically that product-formula approaches can perform better in practice \cite{Childs2018}. 
Furthermore, the LCU method uses ancilla qubits and involves implementing both an operation that coherently performs the constituent unitaries conditioned on the ancilla and a reflection about a certain ancilla state.
Our approach uses no ancillas and only involves evolution according to terms of the Hamiltonian, as it directly implements the time evolution using product formulas, achieving a gate complexity of $O(\alpha T(\alpha T/\epsilon)^{1/(k-1)})$ for arbitrary fixed $k$.

Reference \cite{Haah_2021} uses Lieb-Robinson bounds to create a protocol for quantum simulation of lattice models that resembles the \THR algorithm described in Eq.~(\ref{eq:thrift_1}), but where the splitting of the Hamiltonian is decided based on the support of its summands, not on the energy scales involved in the Hamiltonian. 
The cost of this method is nearly optimal as a function of system size as well as evolution time and approximation error.
However, in practice, this strategy may perform worse than straightforward application of product formulas~\cite{Childs_2019}.

After completing an initial version of this work, we became aware of Ref.~\cite{Omelyan_2002_optimized}. There, \citeauthor{Omelyan_2002_optimized} derive a set of optimised fourth-order product formulas for a Hamiltonian $H = H_A + H_B$ by adding additional sub-steps to standard Trotter formulas and numerically minimizing the error arising from commutators (see \cref{app:subsubsec:omelyan} for details). The generalisation to Hamiltonians with an arbitrary number of terms is described in Ref.~\cite{Ostmeyer_2023_trotter}. Of particular interest for the present work is an optimised formula valid in the regime $\alpha \ll 1$ of a Hamiltonian $H = H_0 + \alpha H_1$ (denominated ``Omelyan's small $A$'' in Ref.~\cite{Ostmeyer_2023_trotter}). As we discuss in \cref{sec:numerical_results}, for the 1D and 2D transverse-field Ising and the 1D Heisenberg models, \THR outperforms this optimised formula for all the values of $\alpha$ we consider. On the other hand, Omelyan et al.'s optmised small $A$ formula proves to be the most efficient algorithm in the $\hop/U \ll 1$ regime of the \FH model. This is mainly due to the high cost of implementing the terms arising in the \THR decomposition of this model. 

\section{Motivation and main result}\label{sec:motivation}

Consider a Hamiltonian of the form $H=H_{0}+\alpha H_{1}$
where $\alpha\ll1$, the norms of $H_0$ and $H_1$ are comparable, and the unitary $U_0=e^{-itH_0}$ can be implemented exactly for arbitrary times $t$ with an efficient quantum circuit, with complexity independent of $t$.
We are interested in approximating the full evolution operator $U=e^{-itH}$. The first-order Trotter formula with $N$ steps
has error \cite{Childs_2019, Ostmeyer_2023_trotter}
\begin{equation}
\|e^{-it(H_{0}+\alpha H_{1})}-(e^{-i\frac{t}{N}H_{0}}e^{-i\frac{t}{N}\alpha H_{1}})^N\| \leq \frac{t^2|\alpha|}{2N}\|[H_{0},H_{1}]\|.
\end{equation}

We can use the fact that $U_0$ is implementable exactly to give a simulation with lower error. Going to the interaction (also known as intermediate) picture~\cite{fetter2003quantum}, 
we have 
\begin{align}\label{inter_pic}
U & =\lim_{N\rightarrow\infty}\prod_{k=1}^{N}e^{-i\frac{t}{N}H_{0}}e^{-i\frac{t}{N}\alpha H_{1}},\nonumber \\
 &=e^{-itH_0}\lim_{N\rightarrow\infty}e^{\frac{i(N-1)t}{N}H_{0}}e^{-i\frac{t}{N}\alpha H_{1}}e^{-i\frac{(N-1)t}{N}H_{0}}\dots e^{-i\frac{t}{N} \alpha H_{1}}e^{i\frac{t}{N}H_{0}}e^{-i\frac{t}{N}\alpha H_{1}}e^{-i\frac{t}{N}H_{0}}e^{-i\frac{t}{N}\alpha H_{1}},\\
 & =e^{-itH_0}\mathcal{T}e^{-i\int_{0}^{t}\alpha H_{1}(\tau)d\tau},\nonumber 
\end{align}
where in the second line we have just inserted identities between
each exponential of $H_{1}$. Here, $\mathcal{T}$ is the time-ordering
operator (which moves terms with smaller times to the right) and $H_{1}(t)=e^{itH_{0}}H_{1}e^{-itH_{0}}$. This is a better
starting expression for bounding the error.
Let $[\mathcal{T}e^{-i\int_{0}^{t}\alpha H_{1}(\tau)d\tau}]_{\rm \apx}$ 
denote a product formula (to be defined) for approximating $\mathcal{T}e^{-i\int_{0}^{t}\alpha H_{1}(\tau)d\tau}$, and let $U_{\rm \apx}$ denote the overall approximation to $U$ obtained by using this formula.
Then we have
\begin{align}
\|U-U_{\rm \apx}\| & =\|e^{-itH_{0}}\mathcal{T}e^{-i\int_{0}^{t}\alpha H_{1}(\tau)d\tau}-e^{-itH_{0}}[\mathcal{T}e^{-i\int_{0}^{t}\alpha H_{1}(\tau)d\tau}]_{\rm \apx}\|\nonumber \\
 & =\|\mathcal{T}e^{-i\int_{0}^{t}\alpha H_{1}(\tau)d\tau}-[\mathcal{T}e^{-i\int_{0}^{t}\alpha H_{1}(\tau)d\tau}]_{\rm \apx}\|\label{eq:time_dep_trot}
\end{align}
by invariance of the operator norm under unitary transformations.
Using for example
the first-order generalised Trotter formula $[\mathcal{T}e^{-i\int_{0}^{t}\alpha H_{1}(\tau)d\tau}]_{\rm \apx}=\mathcal{T}e^{-i\int_{0}^{t}\alpha H_{1}^{A}(\tau)d\tau}\mathcal{T}e^{-i\int_{0}^{t}\alpha H_{1}^{B}(\tau)d\tau}$ \cite{J_Huyghebaert_1990,Poulin_2011},
where $H_{1}(\tau)=H_{1}^{A}(\tau)+H_{1}^{B}(\tau)$ is some splitting of $H_1(\tau)$, we have
\begin{align}
\|U-U_{\rm \apx}\| & =\|\mathcal{T}e^{-i\int_{0}^{t}\alpha H_{1}(\tau)d\tau}-\mathcal{T}e^{-i\int_{0}^{t}\alpha H_{1}^{A}(\tau)d\tau}\mathcal{T}e^{-i\int_{0}^{t}\alpha H_{1}^{B}(\tau)d\tau}\|,\nonumber \\
 & \leq\alpha^{2}\int_{0}^{t}dv\int_{0}^{v}ds\|[H_{1}^{A}(s),H_{1}^{B}(v)]\|=O(\alpha^2t^2),
 \label{eq:thrift_error}
\end{align}
assuming that $\|[H_{1}^{A}(s),H_{1}^{B}(v)]\| = O(1)$. Note that the error now scales as $\alpha^{2}$ instead of $\alpha$. For general evolution time, we can divide the evolution into $N$ steps, giving an error
\begin{align}\nonumber
\|U-U_{\rm \apx}\| & =\|\mathcal{T}e^{-i\int_{0}^{t}\alpha H_{1}(\tau)d\tau}-\prod_{j=0}^{N-1}\mathcal{T}e^{-i\int_{j\frac{t}{N}}^{(j+1)\frac{t}{N}}\alpha H_{1}^{A}(\tau)d\tau}\mathcal{T}e^{-i\int_{j\frac{t}{N}}^{(j+1)\frac{t}{N}}\alpha H_{1}^{B}(\tau)d\tau}\|\\
 & \leq\alpha^{2}\sum_{j=0}^{N-1}\int_{j\frac{t}{N}}^{(j+1)\frac{t}{N}}dv\int_{j\frac{t}{N}}^{v}ds\|[H_{1}^{A}(s),H_{1}^{B}(v)]\|=O\left(\frac{\alpha^2t^2}{N}\right).
\end{align}

To turn this approach into a useful product-formula decomposition, we describe how to implement the time-ordered exponentials. This can be done using the definition of the time-ordered exponential in the other direction,
\begin{align}\label{eq:time_ord_simp}
 \mathcal{T}e^{-i\int_{a}^{b}d\tau\alpha A(\tau)}& =e^{ibH_{0}}e^{-i(b-a)(H_{0}+\alpha A)}e^{-iaH_{0}},
\end{align}
which is valid for any Hermitian operator $A(t)=e^{iH_{0}t}Ae^{-iH_{0}t}$. This leads to the decomposition
\begin{align}\label{eq:thrift_1}
U_{\rm \apx}=e^{-itH_0}\mathcal{T}e^{-i\int_0^t\alpha H_1^A(\tau)d\tau}\mathcal{T}e^{-i\int_0^t\alpha H_1^B(\tau)d\tau}=e^{-it(H_0+\alpha H_1^A)}e^{itH_0}e^{-it(H_0+\alpha H_1^B)}.
\end{align}
This is nothing more than the usual first-order Trotter decomposition of the Hamiltonian $H=H_0+\alpha(H_1^A+H_1^B)$ using the summands $H_0+\alpha H_1^A$, $-H_0$, and $H_0+\alpha H_1^B$.

The decomposition (\ref{eq:thrift_1}) has an error $\alpha$ times smaller than the usual first-order Trotter formula. In particular, we have the following theorem.

\begin{theorem}[\THR decomposition] \label{thm:thrift}
  Given a Hamiltonian $H=H_0+\alpha H_1$ where $H_1=\sum_{\gamma=1}^\Gamma H_1^\gamma$,
 the decomposition
     \begin{align}\label{eq:Thrift_simple}
 U_{\rm \apx}(t):=e^{-itH_0}\prod_\gamma \left(e^{itH_0}e^{-it(H_0+\alpha H_1^\gamma)}\right)
 \end{align}
approximates $U(t)=e^{-itH}$ with error 
\begin{equation}\label{eq:thrift_error_proof}
\|U(t)-U_{\rm apx}(t)\|\leq \alpha^2\int_{0}^{t}dv\int_{0}^{v}ds\sum_{\gamma_1<\gamma_2=1}^\Gamma\|[H_{1}^{\gamma_1}(s),H_{1}^{\gamma_2}(v)]\|.
\end{equation}
For sufficiently small time, this error is $O(\alpha^2t^2)$.
\end{theorem}

\begin{proof}
Define the approximant 
\begin{align}
V^{(j)}(t):=\left (\prod_{k=1}^j\mathcal{T}e^{-i\int_0^tH_1^k(s)ds}\right)\mathcal{T}e^{-i\int_0^t\sum_{k=j+1}^\Gamma H_1^k(s)ds}.
\end{align}
Here $V^{(0)}(t)=\mathcal{T}e^{-i\int_0^tH_1(s)ds}$ corresponds to the evolution under the full Hamiltonian $H_1(t)$, while $V^{(\Gamma-1)}(t)=e^{itH_0}U_{\rm apx}(t)$, where $U_{\rm apx}(t)$ is defined in \cref{eq:Thrift_simple}. This follows from repeated use of \cref{eq:time_ord_simp}. Using the invariance of the operator norm and \cref{eq:thrift_error}, it follows that
\begin{align}\label{eq:proof_thrift}
\|V^{(j)}(t)-V^{(j+1)}(t)\|\leq \alpha^2\int_{0}^{t}dv\int_{0}^{v}ds\sum_{k=j+2}^\Gamma\|[H_{1}^{j+1}(s),H_{1}^{k}(v)]\|.
\end{align}
We use \cref{eq:proof_thrift} to bound the error by applying the triangle inequality on the identity $V^{(0)}-V^{(\Gamma-1)}=\sum_{j=0}^{\Gamma-2} (V^{(j)}-V^{(j+1)})$ and noting that $\| V^{(0)}(t)-V^{(\Gamma-1)}(t)\|=\|U(t)-U_{\rm apx}(t) \|$, which leads finally to
\cref{eq:thrift_error_proof} as claimed.
\end{proof}

For $\alpha$ small 
the error of this approximation scales better than a normal Trotter approximation.

The \THR decomposition in \cref{thm:thrift} corresponds to a first-order Trotter formula, and can be used as a seed for higher-order approximations using standard techniques \cite{Berry_2006,Childs2021,Susuki1991,morales2022greatly}.
More formally, we have the following procedure to turn a product formula into a \THR formula with $O(\alpha^2)$ error scaling.

\begin{proposition}[Higher-order \THR] \label{thm:higher_order_thrift}
  Given a second-order product formula $\mathcal{S}_2(t)$ and a set of parameters $\{u_j\}_{j=1}^m$ such that 
  \begin{equation}
    \mathcal{S}_k(t) = \prod_{j=1}^m \mathcal{S}_2(u_j t)
    \label{eq:product_formula}
  \end{equation}
  is a $k$th-order product formula, the product 
  \begin{equation}
  \mathcal{S}_k(t) = \prod_{j=1}^m U_{\rm \apx}\left(\frac{u_j}{2} t\right) U^\dagger_{\rm \apx}\left(-\frac{u_j}{2} t\right),
    \label{eq:symmetric_thrift}
  \end{equation}
with $U_{\rm \apx}(t)$ specified by \cref{eq:Thrift_simple},
  approximates $e^{-itH}$ with error $O(t^{k+1} \alpha^2)$.
\end{proposition}

\begin{proof}
  $U_{\rm \apx}(t)$ is simply a first-order product formula with the 
  unusual splitting
  \begin{equation}
    H = (H_0 + \alpha H_1^1) - H_0 + \cdots + (H_0 + \alpha H_1^\Gamma).
  \end{equation}
  It follows trivially that \cref{eq:symmetric_thrift} is a $k$th-order product formula.
  To prove the $O(\alpha^2)$ error scaling, we write
  \begin{equation}
  \mathcal{S}_k(t)=e^{-i\sum_{j}u_{j}tH_{0}}\prod_{j=1}^{m}\left(\prod_{\gamma=1}^{\Gamma}\mathcal{T}e^{-i\int_{(a_{j}+\frac{u_{j}}{2})t}^{(a_{j}+u_{j})t}\alpha H_{1}^{\gamma}(\tau)d\tau}\right)\left(\prod_{\gamma=\Gamma}^{1}\mathcal{T}e^{-i\int_{a_{j}t}^{(a_{j}+\frac{u_{j}}{2})t}\alpha H_{1}^{\gamma}(\tau)d\tau}\right)
  \end{equation}
  with $a_{m-k}=\sum_{r=0}^{k-1}u_{m-r}$ and $a_m=0$. This expression follows by applying  \cref{eq:time_ord_simp} to \cref{eq:symmetric_thrift}.
  A valid product formula satisfies $\sum_{j=1}^mu_j=1$, so the claim follows by \cref{thm: alpha squared scaling general} of \cref{appendix_2}.
  \end{proof}

\section{Beyond quadratic scaling}

The procedure developed in \cref{thm:higher_order_thrift} improves the $O(t)$ error scaling, but leaves the $O(\alpha^2)$ error scaling unchanged. In fact, in \cref{appendix_2} we prove that no formula that approximates the evolution by a product of time-ordered evolutions according to terms of the Hamiltonian can achieve better scaling in $\alpha$ than \THR, regardless of how the Hamiltonian is decomposed. However, in this section we show how to achieve better scaling using two alternative approaches.

Motivated by \cref{inter_pic}, we look for approximations of the time-ordered operator that have better error scaling in the small parameter $\alpha$. 
First, we consider the Magnus expansion \cite{Magnus1954}, which approximates the time-ordered exponential as the standard exponential of a time-dependent operator $\Omega$.
Second, we consider directly approximating the time-ordered exponential as a product of exponentials \cite{Fer1958}. 
We show that these approaches achieve error scaling $O(t^k\alpha^k)$ for any positive integer $k$. We also present two algorithms to implement these approximations in practice.

\subsection{Magnus-\THR}\label{sec:magnus}

Writing
\begin{align}
    \mathcal{T}e^{-i\alpha\int_0^tH(s)ds}=:e^{\Omega(\alpha,t)}
\end{align}
for some time-dependent operator $\Omega(\alpha,t)$,
 it is easy to show that $\frac{de^{\Omega(t)}}{dt}e^{-\Omega(t)}=-i\alpha H_1(t)$. Magnus \cite{Magnus1954} used this to find an equation for $\Omega$ by employing the inverse of the derivative of the exponential map, i.e.,
\begin{align}\label{eq:Magnus_map}
    \frac{de^{\Omega(t)}}{dt}e^{-\Omega(t)}=\frac{e^{\ad_\Omega}-1}{\ad_\Omega}\frac{d\Omega}{dt} \quad\rightarrow\quad \frac{d\Omega}{dt}=\frac{\ad_\Omega}{e^{\ad_\Omega}-1}(-i\alpha H_1)=\sum_{k=0}^\infty\frac{b_k}{k!}\ad^k_\Omega(-i\alpha H_1),
\end{align}
where $\ad_\Omega(\cdot):=[\Omega,\cdot]$ and $\ad_\Omega^j(\cdot):=\ad_\Omega^{j-1}([\Omega,\cdot])$. The coeficients $b_j$ are Bernoulli numbers, defined through $\frac{x}{e^x-1}=\sum_{j=0}^\infty\frac{b_j}{j!}x^j$.
The equation for $\Omega$ can now be solved through Picard iteration \cite{Magnus1954,BLANES2009}. Defining $\alpha$-independent coefficients $\tilde{\Omega}_j(t)$ so that  $e^{\Omega(\alpha,t)}=\exp\left({\sum_{j=1}^\infty\alpha^j\tilde{\Omega}_j(t)}\right)$, and using this expression in \cref{eq:Magnus_map}, produces the recurrence \cite{Klarsfeld1989}
\begin{align}
\label{eq:tilde_omega}
    \frac{d}{dt}\tilde{\Omega}_{n}(t)&=\sum_{k=1}^{n-1}\frac{b_{k}}{k!}\sum_{\substack{j_{1}+j_{2}+\dots+j_{k}=n-1\\j_1,j_2,\dots,j_k\geq 1}}[\tilde{\Omega}_{j_{1}}(t),[\tilde{\Omega}_{j_{2}}(t),\dots[\tilde{\Omega}_{j_{k}}(t),-iH_{1}(t)]\dots]].
\end{align}

The series for $\Omega$ converges for sufficiently small time $t$ \cite{Blanes1998,moan1998efficient} (see also \cref{app:theorem_conv}).
Using these results, we can state the following lemma bounding the terms of the Magnus expansion.

\begin{lemma}\label{lem:bound_omega}
For $l\geq 1$,
$\|\tilde{\Omega}_l(t)\|\leq \frac{1}{2}x_l(2\int_0^t\|H_1(s)\| ds)^l$, where $x_l$ is the coefficient of $s^l$ in the expansion of $G^{-1}(s)=\sum_{m=1}^\infty x_m s^m$, the inverse function of $G(s)=\int_0^s(2+\frac{x}{2}(1-\cot(x/2))^{-1}dx$.
\end{lemma}

This lemma is mentioned in \cite{BLANES2009}. We include a proof for completeness in \cref{app:convergence}. Armed with \cref{lem:bound_omega}, we can now easily prove the following approximation theorem.

\begin{theorem}[Magnus-\THR decomposition] \label{thm:magnus_thrift}
Consider a Hamiltonian $H=H_0+\alpha H_1$. Let $H_1(t):=e^{itH_0}H_1e^{-itH_0}$. Defining $\Omega^{[k]}:=\sum_{j=1}^k\Omega_j(\alpha,t)$, the operation
\begin{align}\label{eq:approx_Magnus}
U_{M}(t):=e^{-itH_0}\exp\left(\Omega^{[k]}(\alpha,t)\right)
=e^{-itH_0}\exp\left(\sum_{j=1}^k\alpha^j\tilde{\Omega}_j(t)\right)
\end{align}
approximates $U(t)=e^{-itH}$ with error $O((t\alpha)^{k+1})$ for small times $t$.
\end{theorem}

\begin{proof}
As $e^{-it(H_0+\alpha H_1)}=e^{-itH_0}\mathcal{T}e^{-i\alpha\int_0^tH_1(s)}$, it suffices to approximate the time-ordered evolution $\mathcal {U}(\alpha,t):=\mathcal{T}e^{-i\alpha\int_0^tH_1(s)}$. 
 Introducing the Taylor remainder of a function $h(\alpha)$ as $R_k(h(\alpha)):=\sum_{n=k+1}^\infty\frac{\alpha^n}{n!}h^{(n)}(0)$,  
it follows that for $\Omega(\alpha,t)=\sum_{j=1}^\infty\alpha^j\tilde{\Omega}_j(t)$,
\begin{align}\nonumber
    \|R_k(\Omega(\alpha,t))\|&\leq\sum_{n=k+1}^\infty\alpha^n\|\tilde{\Omega}_n(t)\|
    \quad\mbox{using the triangle inequality and the definition of the remainder}
    \\\nonumber
    &\leq \frac{1}{2}\sum_{n=k+1}^\infty\frac{\alpha^n}{n!}\frac{d^{n}}{dz^{n}}(G^{-1}(0))\left(2\int_{0}^{t}\|H_1(x)\|dx\right)^{n}\quad\mbox{applying \cref{lem:bound_omega} termwise}\\\label{eq:remainder_bound}
    &=R_{k}\left(\frac{1}{2}G^{-1}\left(2\alpha\int_0^t \|H_1(s)\|ds\right)\right)
    \quad\mbox{using the definition of the remainder}.
\end{align}

The remainder provides a bound on the difference between  $\mathcal{U}(\alpha,t)=e^{\Omega(\alpha,t)}=e^{(\Omega^{[k]}(\alpha,t)+R_{k}(\Omega(\alpha,t))}$ and $e^{\Omega^{[k]}(\alpha,t)}$ by means of the integral representation of the error
\begin{align}
F:=e^{\Omega(\alpha,t)}e^{-\Omega^{[k]}(\alpha,t)}-1=\int_0^sds e^{s(\Omega^{[k]}(\alpha,t)+R_{k}(\alpha,t))}
R_{k}(\alpha,t)e^{-s\Omega^{[k]}(\alpha,t)}.
\end{align}
Using \cref{eq:remainder_bound}, we have $\|\mathcal{U}(\alpha,t)-e^{\Omega^{[k]}}\|\leq R_k(\frac{1}{2}G^{-1}(\alpha t \|H_1\|))$. This implies that the error scales as $O((\alpha t)^{k+1})$.
\end{proof}

Note that the above proof extends trivially to an arbitrary time-dependent $H_1(t)$.

\subsubsection*{Magnus-\THR Algorithm}

We now describe a method for approximating the dynamics of the Hamiltonian $H=H_{0}+\alpha H_{1}$ for time $T$ using the Magnus expansion.
The approach is as follows:
\begin{enumerate}
\item Write the evolution operator $U(T)=e^{-iT(H_{0}+\alpha H_{1})}$
in the interaction picture, with $H_{0}$ as the dominant part:
\begin{align}
\label{eq:U_interaction_picture}
U(T)=e^{-iTH_{0}}\mathcal{T}e^{-i\int_{0}^{T}\alpha H_{1}(t)}.
\end{align}
\item Slice the time $T$ into $N$ intervals:
\begin{align}
\label{eq:Texp_interaction_picture}
\mathcal{T}e^{-i\int_{0}^{T}\alpha H_{1}(t)}=\prod_{k=0}^{N-1}\mathcal{T}e^{-i\int_{k\frac{T}{N}}^{(k+1)\frac{T}{N}}\alpha H_{1}(t)}.
\end{align}
\item Approximate the time-ordered exponential of a slice using its Magnus
expansion up to order $O((\frac{T}{N}\alpha)^{p})$. Note that here we use the Magnus expansion with an initial time $t_0\neq 0$. We write the Magnus approximation of order $p$ with an arbitrary initial time $t$  as $\Omega(\alpha,\delta t ;t)$, such that
\begin{align}
\mathcal{T}e^{-i\int_{t}^{t+\delta t}\alpha H_{1}(t)}=\exp\left(\Omega^{[p]}(\alpha,\delta t;t)\right)+O((\delta t\alpha)^{p+1}).
\end{align}
\item Approximate the exponential $\exp\left(\Omega^{[p]}(\alpha,\delta t;t)\right)$
obtained from the Magnus expansion using a $p$th-order product formula $S_p$:
\begin{align}
\exp\left(\Omega^{[p]}(\alpha,\delta t;t)\right)=S_{p}(t,\delta t)+O((\delta t\alpha)^{p+1}). 
\end{align}
This procedure leads to the decomposition
\begin{align}\label{eq:decomp magnus}
U(T)=e^{-iTH_{0}}\prod_{k=1}^{N}S_{p}\left((k-1)\frac{T}{N},\frac{T}{N}\right)+O\left(N\left(\frac{T\alpha}{N}\right)^{p+1}\right).
\end{align}
\end{enumerate}

As an example, consider the expansion of 
\begin{align}
e^{\Omega^{[2]}(\alpha,t;\delta t)}	=e^{-i\alpha\delta t\left(\frac{1}{\delta t}\int_{t}^{t+\delta t}d\tau H(\tau)-\frac{i\alpha}{2\delta t}\int_{t}^{t+\delta t}dt_{1}\int_{t}^{t_{1}}dt_{2}[H(t_{1}),H(t_{2})]\right)}.
\end{align}
Expanding the time-dependent Hamiltonian as a sum of time-independent operators $O_{q}$ and functions of time $\alpha_{q}(t)$ as $H(t)=\sum_{q=1}^{Q}\alpha_{q}(t)O_{q}$, we find
\begin{align}
    \label{eq:Magnus_2_implementation}
    \Omega^{[2]}(t,\delta t)=	-i\alpha\delta t\left(\sum_{q}A_{q}(t,\delta t)O_{q}+\sum_{q>p}B_{qp}(t,\delta t)[O_{q},O_{p}]\right)
\end{align}
where 
\begin{align}
    A_{q}(t,\delta t)
    &=\frac{1}{\delta t}\int_{t}^{t+\delta t}d\tau\alpha_{q}(\tau), \\
    B_{qp}(t,\delta t)
    &=-\frac{i\alpha}{4\delta t}\int_{t}^{t+\delta t}\int_{t}^{t+\delta t}dt_{1}dt_{2}\alpha_{q}(t_{1})\alpha_{p}(t_{2}){\rm sign}(t_{1}-t_{2}),
\end{align}
which can be computed classically. Thus we can approximate $e^{\Omega^{[2]}(\alpha,t;\delta t)}$ using a second-order product formula as
\begin{align}\label{eq:Magnus_2_example}
    e^{\Omega^{[2]}(\alpha,t;\delta t)}	&=e^{-i\epsilon\delta t\left(\frac{1}{\delta t}\int_{t}^{t+\delta t}d\tau H(\tau)-\frac{i\epsilon}{2\delta t}\int_{t}^{t+\delta t}dt_{1}\int_{t}^{t_{1}}dt_{2}[H(t_{1}),H(t_{2})]\right)}\nonumber\\
	&=e^{-i\epsilon\delta t\left(\sum_{q}A_{q}(t,\delta t)O_{q}+\sum_{q>p}B_{qp}(t,\delta t)[O_{q},O_{p}]\right)}\nonumber\\
	&=e^{-i\frac{\epsilon\delta t}{2}\sum_{q}A_{q}(t,\delta t)O_{q}}e^{-i\epsilon\delta t\sum_{q>p}B_{qp}(t,\delta t)[O_{q},O_{p}]}e^{-i\frac{\epsilon\delta t}{2}\sum_{q}A_{q}(t,\delta t)O_{q}}+O(\alpha^{3}\delta t^{3}).
\end{align}
If necessary, each of the products can be decomposed further using a second-order product formula to keep the error at most $O(\alpha^{3}\delta t^{3})$.

Note that in any application of these formulas, some care has to be taken when expanding functions of time, to avoid losing the favourable scaling with $\alpha$. As the error scales with both $\alpha$ and $t$, in any expansion the scaling with both of them should be considered.

In principle it should be possible to analyse the commutator scaling of the product formula appearing in \cref{eq:decomp magnus}, generalizing \cite{Childs2021}. We leave this as a topic for future work.

\subsection{Fer-\THR} \label{sec:fer}

We can bypass approximating the Magnus term $e^{\Omega^{[j]}(\alpha, t:\delta t)}$ in \cref{eq:approx_Magnus} by directly looking for an approximation of the time-ordered operator as a product of exponentials. This approach generates the following decomposition.

As before, the starting point is an approximation of the time-ordered operator in the interaction picture. For this approximation, Fer \cite{Fer1958} postulated the form
\begin{align}
\mathcal{T}e^{-i\int_0^tA(s)ds}=e^{-i\int_0^tA(s)ds}V(t).
\end{align}
This implies the equation
\begin{align}
\frac{d}{dt}V=\left[-ie^{i\int_0^tA(s)ds}A(t)e^{-i\int_0^tA(s)ds}-e^{i\int_0^tA(s)ds}\frac{d}{dt}e^{-i\int_0^tA(s)ds}\right]V=:-iA_1(t)V,
\end{align}
which can be formally solved as $V=\mathcal{T}e^{-i\int_0^tA_1(s)ds}$. Repeating this procedure $k$ times gives
\begin{align}
\mathcal{T}e^{-i\int_0^tA(s)}=\prod_{j=0}^{k-1}e^{-i\int_0^tA_j(s)ds}V_{k},
\end{align}
where $A_0:=A$ and
\begin{align}\nonumber\label{eq:rec_Fer}
    A_j(t)&=e^{i\int_0^tA_{j-1}(s)ds}A_{j-1}(t)e^{-i\int_0^tA_{j-1}(s)ds}-ie^{i\int_0^tA_{j-1}(s)ds}\frac{d}{dt}e^{-i\int_0^tA_{j-1}(s)}ds\\
    &=\sum_{m=1}^\infty (-1)^m\frac{ m}{(m+1)!}\mathrm{ad}^m_{-i\int_0^tA_{j-1}(s)ds}(A_{j-1}(t)).
\end{align}
Setting $V_k=1$ truncates this product, giving an approximation of order $O(t^{2^{k+1}-1})$ \cite{Iserles1984}.

This analysis can be modified slightly to determine how the error depends on a scaling factor $\alpha$ by making the substitution $A_0\rightarrow \alpha A_0$. For the following we absorb the factor of $-i$ into $A_0$ as it does not change the analysis.
\begin{lemma}\label{lem: Ak scaling}
    Let $A_0(t)$ be an operator-valued function that is analytic in $t$ over the reals. 
    For a real scaling factor $\alpha$, define $\alpha A_k(t)$ recursively as
    \begin{equation}
        \alpha A_{k+1}(t) = 
        \sum_{m=1}^\infty (-1)^m \frac{m}{(m+1)!}
        \ad^m_{\int_0^{t}\alpha A_k(s)ds}[\alpha A_k(t)].
    \end{equation}
    If $\alpha A_k(t)=O(\alpha^qt^p)$ then $\alpha A_{k+1}(t)=O(\alpha^{2q}t^{2p+2})$.
\end{lemma}

\begin{proof}
    The proof is largely similar to the proof of Lemma 2 of \cite{Iserles1984}, differing in the fact that it also tracks the scaling variable $\alpha$. For notational compactness, let $\alpha B_k(t) = \int_0^{t}\alpha A_k(s)ds$.

    By Lemma 1 of \cite{Iserles1984}, $\alpha A_k$ is analytic in $t$ over the reals for all $k$. 
    As $A_k(t)$ has no dependence on $\alpha$, they are also analytic over all
    $\alpha$. We may then write
    \begin{equation}
    \begin{split}
        \alpha A_k(t) =& \sum_{i=0}^{2p+1} \frac{1}{i!}\alpha A_k^{(i)}(0)t^i + 
        \alpha^qt^{2p+2}E_A(\alpha, t), \\
        \alpha B_k(t) =& \sum_{i=1}^{2p+1} \frac{1}{i!}\alpha A_k^{(i-1)}(0)t^i + 
        \alpha^qt^{2p+2}E_B(\alpha, t), \\
    \end{split}
    \end{equation}
    where the superscripts of $A_k$ denote derivatives with respect to $t$.  

    By the bilinearity of the commutator, we have
    \begin{equation}
        \begin{split}
        [\alpha B_k(t), \alpha A_k(t)] =
        \sum_{i=1}^\infty\sum_{j=0}^\infty
        \frac{1}{i!j!} [\alpha A_k^{(i-1)}(0),\alpha A_k^{(j)}(0)]t^{i+j}+
        \alpha^{2q}t^{2p+2}E_1(\alpha, t)
        \end{split}
    \end{equation}
    where we have used the fact that $\alpha A^{(i)}_k(0)\in O(\alpha^q)$. Reordering the summation gives 
    \begin{equation}
        \begin{split}
        [\alpha B_k(t), \alpha A_k(t)] =
        \sum_{i=1}^\infty\frac{1}{i!}\left[\sum_{j=0}^j\binom{i}{j}
        [\alpha A_k^{(i-j-1)}(0),\alpha A_k^{(j)}(0)]\right]t^{i}+
        \alpha^{2q}t^{2p+2}E_2(\alpha, t).
        \end{split}
    \end{equation}

    As $\alpha A_k(t)=O(t^p)$, $\alpha A^{(j)}_k(0)=0$ for $0\leq j \leq p-1$, so for $i\leq 2p$,
    \begin{equation}
        \sum_{j=0}^j\binom{i}{j}[\alpha A_k^{(i-j-1)}(0),\alpha A_k^{(j)}(0)]=0
    \end{equation}
    and
    \begin{equation}
        \sum_{j=0}^{2p+1}\binom{2p+1}{j}
        [\alpha A_k^{(2p-j)}(0),\alpha A_k^{(j)}(0)]=
        \binom{2p+1}{p}[\alpha A_k^{(p)}(0), \alpha A_k^{(p)}(0)]=0.
    \end{equation}
    Therefore $[\alpha B_k(t),\alpha A_k(t)]=O(\alpha^{2q}t^{2p+2})$.
    For the nested commutators we have
    \begin{equation}
        \ad_{\alpha B_k(t)}^m [\alpha A_k(t)] = O(\alpha^{mk}t^{m(p+1)})
    \end{equation}
    because $\alpha B_k(t) = O(\alpha^qt^{p+1})$, so $\alpha A_{k+1}(t)=O(\alpha^{2q} t^{2p+2})$ as claimed.
\end{proof}

\begin{theorem}
     Let
    \begin{equation}
        \begin{split}
        U(t)&=\T e^{\int_0^{t}\alpha A_0(s)ds},\\ 
        U_F(t) &= \prod_{j=0}^{k-1} e^{\int_0^{t}\alpha A_j(s)ds},
        \end{split}
    \end{equation}
    with $A_j$ defined as in \cref{lem: Ak scaling}. Then
    \begin{equation}
        \|U_F(t)-U(t)\| = O(\alpha^{2^{k}}t^{2^{k+1}-1}).
    \end{equation}
\end{theorem}
\begin{proof}
   The proof
   of Theorem 3 of \cite{Iserles1984} shows that
   \begin{equation}
        U_F(t)-U(t) = -\int_0^t U(t-\tau)U_F(\tau)\alpha A_{k}(\tau).
   \end{equation}
   The bound follows since $U(t),U_F(t)=O(1)$ and $\alpha A_{k}(t)=O(\alpha^{2^{k}}t^{2^{k+1}-2})$.
\end{proof}

For an approximation of the total evolution in the interaction picture, we have the following.

\begin{corollary}[Fer-\THR decomposition]
Consider a Hamiltonian $H=H_0+\alpha H_1$, and let $H_1(t)=e^{itH_0}H_1e^{-itH_0}$. Define  
\begin{align}
U_{F}(t)=e^{-itH_0}\prod_{j=0}^{k-1}e^{-i\int_0^tA_j(s)ds},
\end{align}
where $A_j(t)$ is defined recursively from \cref{eq:rec_Fer} with $A_0(t):=\alpha H_1(t)$. Then $U_F(t)$ approximates $U(t)=e^{-itH}$ up to $O(\alpha^{2^k}t^{2^{k+1}-1})$ for small times $t$.
\end{corollary}

Note that the surprising scaling of this approach with $t$ and $\alpha$ is due to the assumption that the unitaries $e^{-i\int_0^tA_j(s)ds}$ can be implemented exactly. In any actual implementation, these unitaries have to approximated up to the target error, thus recovering in practice the same scaling as Magnus-\THR. This is exemplified in the following algorithm.

\subsubsection*{Fer-\THR Algorithm}

To approximate the time evolution generated by the
Hamiltonian $H=H_{0}+\alpha H_{1}$ for time $T$ with precision $O(N(T\alpha/N)^{p+1})$, we perform the following:
\begin{enumerate}
\item Write the evolution operator $U(T)=e^{-iT(H_{0}+\alpha H_{1})}$
in the interaction picture, with $H_{0}$ as the dominant part, i.e.,
\begin{align}
U(T)=e^{-iTH_{0}}\mathcal{T}e^{-i\int_{0}^{T}\alpha H_{1}(t)}.
\end{align}
\item Slice the time $T$ into $N$ intervals:
\begin{align}
\mathcal{T}e^{-i\int_{0}^{T}\alpha H_{1}(t)}=\prod_{k=1}^{N}\mathcal{T}e^{-i\int_{(k-1)\frac{T}{N}}^{k\frac{T}{N}}\alpha H_{1}(t)}.
\end{align}
\item Approximate the time-ordered exponential of a slice using its Fer expansion up to order $O((\frac{T}{N}\alpha)^{p})$:
\begin{align}
\mathcal{T}e^{-i\int_{t}^{t+\delta t}\alpha H_{1}(t)}=\prod_{j=0}^{\log(p)}\exp\left(-i\int_t^{t+\delta t}A_j(s)ds\right)+O((\delta t\alpha)^{p+1}).
\end{align}
\item Approximate each exponential in the product using a $p$th-order formula:
\begin{align}
\exp\left(-i\int_t^{t+\delta t}A_j(s)ds\right)=S^j_{p}(t,\delta t)+O((\delta t\alpha)^{p+1}).
\end{align}
\end{enumerate}

This procedure leads to the decomposition
\begin{align}
U(T)=e^{-iTH_{0}}\prod_{k=1}^{N}\prod_{j=0}^{\log(p)}S^j_{p}\left((k-1)\frac{T}{N},\frac{T}{N}\right)+O\left(N\left(\frac{T\alpha}{N}\right)^{p+1}\right).
\end{align}

Note that for the error in the resulting simulation to have the stated scaling, the unitary $e^{-iTH_0}$ must be implemented with error at most $O((T\alpha)^{p+1})$.

\section{Numerical results} \label{sec:numerical_results}

The asymptotics derived in \cref{thm:thrift,thm:magnus_thrift} show that for
$\alpha$ small enough, \THR methods will outperform Trotter methods, and for even
smaller $\alpha$, Magnus-\THR will eventually outperform \THR.
Similarly, higher-order methods will outperform lower-order methods for small
enough time steps. 
To ascertain that \THR and Magnus-\THR methods give an 
advantage at relevant values of $\alpha$ and $T$, we performed extensive 
simulations of different models, namely the \TFIM in one and 
two dimensions (\cref{sec:tfim}), the Heisenberg model with random local fields in one dimension (\cref{sec:heisenberg}), and the \FH model in one dimension (\cref{sec:fh}).

We compare the ordinary first- and second-order product formulas \cite{Berry_2006,Childs2021}
(here dubbed ``Trotter 1'' and ``Trotter 2''), the fourth-order formula due to 
Suzuki \cite{Susuki1991} (here dubbed ``Trotter 4'' for conciseness), a 
numerically optimised eighth-order product formula due to \textcite{morales2022greatly} 
(``optimised Trotter 8'') based on an ansatz of \textcite{yoshida1990construction}, and a fourth-order formula optimised for Hamiltonians containing a small perturbation derived in Ref.~\cite{Omelyan_2002_optimized} (here dubbed ``opt.\ small $A$ 4'' to indicate its error scaling with $T$). 
For each of these product formulas, we also construct the corresponding \THR
circuit (dubbed ``\THR 1'' through ``\THR 4'' and ``optimised \THR 8'') as described in 
\cref{thm:thrift,thm:higher_order_thrift}.
For the \TFIM, we also implement the Magnus-\THR decompositions described in 
\cref{thm:magnus_thrift} with the first- and second-order Magnus expansion. 

In the numerical implementation of \THR 1 through 8, we use the approximant 
\begin{equation*}
\left(U_{\rm apx}\left({T/N}\right)\right)^N 
= \Bigl(e^{-i\frac{T}{N}H_0}\prod_{\gamma}(e^{i\frac{T}{N}H_0}e^{-i\frac{T}{N}(H_0+\alpha H_1^\gamma)}\Bigr)^N,
\end{equation*}
obtained by first breaking up the total time $T$ into small steps $T/N$ and then approximating each unitary evolution over a small step by \cref{eq:Thrift_simple}. For a total time-independent Hamiltonian $H$, this is equivalent to splitting the time-ordered exponential over the full evolution time into a product of unitary evolutions with a small time step $T/N$, as described in \cref{eq:U_interaction_picture,eq:Texp_interaction_picture}. 

Note that Fer-\THR 1 and Magnus-\THR 1 coincide. As we found that Magnus-\THR 2 was not generally competitive with the other approaches for the systems we analysed, we did not implement Fer-\THR 2 as it has essentially the same cost as Magnus-\THR 2.

\subsection{1D and 2D \TFIM with weak coupling} \label{sec:tfim}

The first model we use for numerical tests and algorithm comparison is the 
\TFIM with weak interaction in one and two dimensions.  In the 1D case, the model is integrable and can be mapped to a free-fermion model that can be simulated 
in polynomial time and space using the method described in \cite{terhal2002flo,Bravyi2012}.
This enables us to simulate chains of length up to $L=100$ using the fermionic linear 
optics simulation tools from \cite{bosse2022floyao}. 
While the equivalence to free fermions makes this model a less interesting target for quantum simulation, we expect that the simulation costs may be indicative of costs for some other 1D models that are not necessarily classically easy. Indeed, we see evidence of this in the case of the Heisenberg model, as shown in \cref{sec:heisenberg}.
In 2D, we are restricted to 
relatively small system sizes using full state vector simulations.

The Hamiltonian of the \TFIM is
\begin{equation}
  \HTFIM = h \sum_j Z_j + J \sum_{\braket{i,j}} X_i X_j,
  \label{eq:tfim_hamiltonian}
\end{equation}
where $X_i$ and $Z_i$ are the spin-$1/2$ operators in the $x$ and $z$ directions,
respectively. For the purpose of studying \THR-based algorithms,
we fix the field strength to $h=1$, let the interaction 
strength $\alpha := J$ be the small parameter, and measure time $T$ in units of $h^{-1}$. Since the transverse-field 
part, $H_0 = \sum_j Z_j$, only consists of one-qubit terms, this has the 
advantage that the interaction-picture Hamiltonian $H_1(t)$ has the same 
locality as the original $H_1 = J \sum_{\braket{i,j}} X_i X_j$, and \THR circuits have the same 2-qubit gate depth as the corresponding Trotter circuits. We also note that, because $e^{-i t J X_i X_j}$
and $e^{-it (J X_i X_j + h (Z_i + Z_j))}$ can be implemented with the same 
number of CNOT gates---namely two---the same holds for CNOT gate depth. The 
2-qubit gate depths of one TDS step for all algorithms considered are shown
in \cref{tab:1d_tfim_step_depth}. The explicit formulas for the approximants used for the \THR simulations of the \TFIM are discussed in \cref{{app:subsubsec:1D_TFIM}}. 

\begin{figure}
\centering
\begin{minipage}[t]{.49\textwidth}
  \centering
  \input{figures/tfim_tds_algo_comparison_1x16.pgf}
  \input{figures/tfim_tds_algo_comparison_3x3.pgf}
  \captionof{figure}{
    (top) Landscape of the best TDS algorithm, as measured by the worst-case error
    $\norm{U - U_{\mathrm{exact}}}$, as a function of the relative field
    strength $\alpha = J/h$ and evolution time $T$ at identical circuit
    depth for a $1 \times 16$ Ising chain with transverse field. The circuit depth was fixed to 1 step of Magnus-\THR 2 evolution. For the other algorithms, the
    number of steps is chosen to match the 2-qubit depth as closely as possible
    according to the 2-qubit depths shown in \cref{tab:1d_tfim_step_depth,tab:2d_tfim_step_depth}.
    The colour of each point represents the algorithm that achieves the lowest
    error at those values of $J$ and $T$, while the brightness indicates
    the magnitude of the error.
    (bottom) Same for a $3 \times 3$ \TFIM. Note that in the top right corner
    of both panels, $\norm{U - U_{\mathrm{exact}}}$ is of order $1$, so this
    region is not of particular interest
  }
  \label{fig:tfim_tds_algo_comparison}
\end{minipage}%
\hfill
\begin{minipage}[t]{.49\textwidth}
  \centering
  \input{figures/tfim_thrift_depth_requirements_J=0.125_worstcase.pgf}
  \input{figures/2dtfim_thrift_depth_requirements_J=0.125_worstcase.pgf}
  \captionof{figure}{
    (top) 2-qubit gate depth to achieve $\norm{U - U_{\mathrm{exact}}} \leq
    0.01$ for the different TDS algorithms for a
    field strength of $J=1/8$ and evolution time $T = L$, for a $1\times L$ Ising chain with transverse field. The 
    depths follow a power law of the form $d = a L^k$ whose parameters $a$ and
    $k$ we determine via a least-squares fit and report, also for different
    values of $\alpha$, in \cref{fig:tfim_tds_algo_fits}.
    (bottom) Similar simulation for a $3 \times 3$ \TFIM. Because the 2D
    \TFIM is not integrable and hence large system sizes are not classically 
    simulable, we fix the system size to $3 \times 3$ and only scale the 
    evolution time $T$. Error bars (mostly barely visible)
    are $\pm 1$, i.e., the minimal possible depth resolution.
  }
  \label{fig:tfim_thrift_depth_requirements_worstcase}
\end{minipage}
\end{figure}

\begin{table}[]
  \centering
  \begin{tabular}{l|rrr}
  Algorithm           & 2-qubit gate depth & CNOT depth  & \# steps in \cref{fig:tfim_tds_algo_comparison} \\ \hline
  Trotter 1           & $2N$               & $4N$        & 15       \\
  Trotter 2           & $2N+1$             & $4N+2$      & 15       \\
  Trotter 4           & $10N+1$            & $20N+2$     & 3        \\
  optimised Trotter 8 & $30N+1$            & $60N+2$     & 1        \\
  \THR 1              & $2N$               & $4N+2$      & 15       \\
  \THR 2              & $2N+1$             & $4N+2$      & 15       \\
  \THR 4              & $10N+1$            & $20N+2$     & 3        \\
  optimised \THR 8    & $30N+1$            & $60N+2$     & 1        \\
  Magnus-\THR 1       & $2N$               & $4N$        & 15        \\
  Magnus-\THR 2       & $12N+3$            & $36N+9$     & 2 \\
  optimised small $A$ 4 & $12N$            & $24N$     & 2    
  \end{tabular}
  \caption{Circuit depth comparison of the different TDS algorithms investigated and shown in \cref{fig:tfim_tds_algo_comparison} for the 1D \TFIM. The first column shows the 2-qubit depth of the circuit corresponding to $N$ Trotter steps in terms of arbitrary 2-qubit gates. The second column shows the corresponding cost in terms of CNOT gates. Finally, the third column gives the number of Trotter steps used in \cref{fig:tfim_tds_algo_comparison}, which correspond to a fixed budget of arbitrary 2-qubit gates of 31.}
  \label{tab:1d_tfim_step_depth}
\end{table}

\begin{table}[]
  \centering
  \begin{tabular}{l|rrr}
  Algorithm           & 2-qubit gate depth & CNOT depth  & \# steps in \cref{fig:tfim_tds_algo_comparison} \\ \hline
  Trotter 1           & $4N$               & $8N$         & 26       \\
  Trotter 2           & $6N+1$             & $12N+2$      & 17       \\
  Trotter 4           & $30N+1$            & $60N+2$      & 3        \\
  optimised Trotter 8 & $90N+1$            & $180N+2$     & 1        \\
  \THR 1              & $4N$               & $8N+2$       & 26       \\
  \THR 2              & $6N+1$             & $12N+2$      & 17       \\
  \THR 4              & $30N+1$            & $60N+2$      & 3        \\
  optimised \THR 8    & $90N+1$            & $180N+2$     & 1        \\
  Magnus-\THR 1       & $4N$               & $8N$         & 26       \\
  Magnus-\THR 2       & $102N+3$           & $306N+9$     & 1 \\
  optimised small $A$ 4 & $28N$            & $56N$     & 3    
  \end{tabular}
  \caption{Circuit depth comparison of the different TDS algorithms investigated and shown in \cref{fig:tfim_tds_algo_comparison} for the 2D \TFIM. The first column shows the 2-qubit depth of the circuit corresponding to $N$ Trotter steps in terms of arbitrary 2-qubit gates. The second column shows the corresponding cost in terms of CNOT gates. Finally, the third column gives the number of Trotter steps used in \cref{fig:tfim_tds_algo_comparison}, which correspond to a fixed budget of arbitrary 2-qubit gates of 105.}
  \label{tab:2d_tfim_step_depth}
\end{table}

In \cref{fig:tfim_tds_algo_comparison} we show which of the different Trotter, 
\THR, or Magnus-\THR algorithms performs best at a given $T$ and $\alpha$
for a wide range of these two quantities for the 1D \TFIM (top) and 2D \TFIM (bottom).
The results broadly agree with what we expect from
\cref{thm:thrift,thm:higher_order_thrift,thm:magnus_thrift}: as $T$ 
decreases, higher-order formulas become advantageous over lower orders, and 
for smaller $\alpha$, \THR methods are advantageous over Trotter methods. Interestingly,
this crossover happens for a relatively large $\alpha \approx 3$ for the \TFIM.
First-order methods are never advantageous for the 1D \TFIM, because for
Hamiltonians that can be split into only two exactly implementable parts 
for Trotterisation, second-order methods have the same amortised depth per step as first-order methods (see \cref{tab:1d_tfim_step_depth}). Magnus-\THR 2 outperforms all other methods only for very small $\alpha < 10^{-2}$ and $T > 1$.

To investigate the scaling of the different algorithms
with the system size and evolution time, we search for the lowest number of 
steps such that each algorithm achieves worst-case error
$\norm{U - U_{\mathrm{exact}}} \leq 0.01$.
For the 1D \TFIM, we scale the system size $L$ and evolution time $T$ 
together as $T = L$. The top plot of \cref{fig:tfim_thrift_depth_requirements_worstcase}
shows the 2-qubit depth to get the error below threshold.
For the 2D \TFIM, we fix the system size at $3 \times 3$ and only change the simulation 
time $T$ when searching for the minimal circuit depth to get 
the error below threshold. The results are shown in the bottom of
\cref{fig:tfim_thrift_depth_requirements_worstcase}. In both cases, we find that 
the circuit depth as a function of evolution time (and system size) is well
described by a power law. The power law exponents match those theoretically 
expected from \cref{sec:system_size_scaling_theory,sec:commutator_scaling}, 
with the notable exception of the optimised eighth-order \THR formula and 
fourth-order Trotter formula, for which the exponents are substantially smaller.
In \cref{sec:additional_numerics_tfim} we show these exponents as
a function of the interaction strength $J = \alpha$ and discuss the results in
more detail. We observe surprisingly slow growth of the  
circuit depth for the optimised eighth-order \THR formula, which appears to scale
sub-linearly in the evolution time. While the small slope of opt.\ \THR 8 for the 1D case could be attributed to the model being fast-forwardable, in the 2D case we believe this is an artifact of the small system size, as the model is not believed to be fast-forwardable in general. See  \cref{sec:additional_numerics_tfim} for more details.
The specific partitions we used to implement Trotter, \citeauthor{Omelyan_2002_optimized}'s, and \THR algorithms for the various models we consider are discussed in \cref{app:sec:circuit_depths_numerical_implementations}.

\subsection{1D Heisenberg model with strong random fields} \label{sec:heisenberg}

The second model we use for numerical tests of the \THR algorithms is the 1D 
spin-$\frac{1}{2}$ Heisenberg model with strong random fields. Unlike the 1D
\TFIM, it is not exactly solvable, and we are not aware of a fast classical simulation 
for arbitrary times. The Hamiltonian is
\begin{equation}
  \HHeisenberg = J \sum_{\braket{i,j}} \left(X_i X_i + Y_i Y_j + Z_i Z_j \right)
                 + \sum_i h_i Z_i,
\end{equation}
where the $h_i$ are chosen uniformly random in $[-h, h]$ and $X_i$, $Y_i$, and $Z_i$ 
are again the spin-$\frac{1}{2}$ operators in the respective directions. We fix $h=1$, use the interaction strength $\alpha:=J$ as the small parameter, and measure time $T$ in units of $h^{-1}$.
To evaluate errors, we always average over 10 different random instantiations 
of the field strengths $h_i$. As in the case of the \TFIM, the field part
$H_0 = \sum_i h_i Z_i$ consists only of one-qubit terms, so 
$H_1(t)$ consists entirely of 2-qubit terms. Because simulating the Heisenberg
interaction $e^{-it (X_i X_j + Y_i Y_j + Z_i Z_j)}$ already takes three 
CNOT gates, simulating the \THR gate $e^{-it (X_i X_j + Y_i Y_j + Z_i Z_j + h_i Z_i + h_j Z_j)}$ takes the same 2-qubit gate depth. Therefore, one step of any \THR circuit takes the same depth as one step of the corresponding Trotter circuit. See \cref{app:subsubsec:1D_Heisenberg} for more details about how we partitioned $\HHeisenberg$. The exact 2-qubit gate depths are shown in \cref{tab:heisenberg_step_depth}. 

\begin{figure}
\centering
\begin{minipage}[t]{.49\textwidth}
  \centering
  \input{figures/heisenberg_tds_algo_comparison_1x8.pgf}
  \captionof{figure}{
    Landscape of the best TDS algorithm, as measured by the worst-case error
    $\Vert U - U_{\mathrm{exact}}\Vert$, as a function of the relative field
    strength $\alpha = J/h$ and evolution time $T$ at identical circuit
    depth for a $1 \times 8$ Heisenberg chain. The circuit depth is
    fixed to one step of optimised \THR 8 evolution. For the other algorithms, the
    number of steps is chosen to match the 2-qubit depth as closely as possible
    according to the 2-qubit depths shown in \cref{tab:heisenberg_step_depth}.
    The colour of each point represents the algorithm that achieves the lowest
    error at those values of $J$ and $T$, while the brightness indicates
    the magnitude of the error. The isolated purple and red pixels in the 
    \THR 4 and \THR 2 regions are artifacts of the randomness in the field
    strengths and running the optimised small $A$ and eighth-order simulations with different
    random fields, but do not seem indicative of the general relative performance of
    the algorithms at these $(\alpha, T)$-points.}
  \label{fig:heisenberg_tds_algo_comparison}
\end{minipage}%
\hfill
\begin{minipage}[t]{.49\textwidth}
  \centering
  \input{figures/heisenberg_thrift_depth_requirements_t=0.125_averagecase.pgf}
  \captionof{figure}{
    2-qubit depth to achieve average infidelity
    $\mathbb{E}_{\{\ket{x}\}}[1 - |\braket{x|U_{\mathrm{exact}}^\dagger U|x}|^2] \leq  0.01$
    for the different TDS algorithms for
    a $1\times L$ Heisenberg chain with field strength of $J = 1/8$ and evolution time $T = L$.
    Unlike \cref{fig:tfim_thrift_depth_requirements_worstcase}, we use
    average fidelity 
    to be able to simulate larger systems.
    Again, the required depths follow a power law of the form $d = a L^k$ whose
    parameters $a$ and $k$ we determine via a least-squares fit and use to
    extrapolate to up to $L = 100$. We report the fit parameters $a$ and $k$,
    also for different values of $\alpha$, in \cref{fig:heisenberg_tds_algo_fits_1d}. Error bars are $\pm 1$ step and the
    shaded regions are the one-sigma confidence intervals of the extrapolations.
  }
  \label{fig:heisenberg_thrift_depth_requirements_averagecase}
\end{minipage}
\end{figure}

\begin{table}[]
    \centering
    \begin{tabular}{l|rrr}
    Algorithm           & 2-qubit gate depth & CNOT depth  & \# steps in      \cref{fig:heisenberg_tds_algo_comparison} \\ \hline
    Trotter 1           & $2N$               & $6N$        & 15       \\
    Trotter 2           & $2N+1$             & $6N+3$      & 15       \\
    Trotter 4           & $10N+1$            & $30N+3$     & 3        \\
    optimised Trotter 8 & $30N+1$            & $90N+3$     & 1        \\
    \THR 1              & $2N$               & $6N+3$      & 15       \\
    \THR 2              & $2N+1$             & $6N+3$      & 15       \\
    \THR 4              & $10N+1$            & $30N+3$     & 3        \\
    optimised \THR 8    & $30N+1$            & $90N+3$     & 1  \\
    optimised small $A$ 4 & $12N$            & $36N$     & 2    
    \end{tabular}
    \caption{
    Circuit depth comparison of the different TDS algorithms investigated and shown in \cref{fig:heisenberg_tds_algo_comparison} for the 1D Heisenberg model. The first column shows the 2-qubit depth of the circuit corresponding to $N$ Trotter steps in terms of arbitrary 2-qubit gates. The second column shows the corresponding cost in terms of CNOT gates. Finally, the third column gives the number of Trotter steps used in \cref{fig:heisenberg_tds_algo_comparison}, which correspond to a fixed budget of arbitrary 2-qubit gates of 31.
    }
    \label{tab:heisenberg_step_depth}
\end{table}

In \cref{fig:heisenberg_tds_algo_comparison,fig:heisenberg_thrift_depth_requirements_averagecase}
we repeat the analysis done for the \TFIM in
\cref{fig:tfim_tds_algo_comparison,fig:tfim_thrift_depth_requirements_worstcase}
for the Heisenberg model. However, because the Heisenberg model is not integrable
and average-case errors are much easier to compute than 
worst-case errors, we use the average infidelity as a figure of merit in 
\cref{fig:heisenberg_thrift_depth_requirements_averagecase}.
(Note that this may not be indicative of worst-case performance, since product formula simulations can have significantly better performance on average \cite{zhao2022hamiltonian}.)
Similarly to the case of the \TFIM, the \THR methods perform better than the 
corresponding Trotter methods, with higher-order methods outperforming 
lower-order methods for smaller $T$ and $\alpha$ in 
\cref{fig:heisenberg_tds_algo_comparison}. We observe that the crossover
point from one method to the next in \cref{fig:heisenberg_tds_algo_comparison}
roughly happens along lines of constant $\alpha T$. This is because the
interaction-picture Hamiltonian $H_1(t)$ scales with $\alpha$, so the
relevant scale for the Trotter errors is $\alpha T/N$. For very
small $\alpha T$, it seems that the optimised eighth-order formula
performs best in \cref{fig:heisenberg_tds_algo_comparison}, but there the
errors are as small as $10^{-15}$, i.e., within the precision typically achieved
by 64-bit floating point computations and much smaller than one could hope
to achieve on real hardware.
In \cref{fig:heisenberg_thrift_depth_requirements_averagecase}
we see that the \THR methods always outperform the corresponding Trotter
methods, and the 2-qubit gate depth to achieve average infidelity below
a fixed threshold scales very similarly with $T$ and the system size $L$
for both methods, in broad agreement with the theory in
\cref{sec:system_size_scaling_theory,sec:commutator_scaling}.  
\Cref{fig:heisenberg_thrift_depth_requirements_averagecase} can also be 
directly compared to Fig. 1 in \cite{zhao2022hamiltonian}, which considers 
the same question (albeit only for Trotter and not for \THR methods) for 
the Heisenberg model at $J=1$. That analysis finds very similar results, including 
matching exponents $k$. We present a more detailed analysis of the
scaling of the circuit depth with system size and evolution time in
\cref{sec:additional_numerics_heisenberg}.

For this model, we did not implement the Magnus-THRIFT algorithm since we expect that it performs similarly to the 1D \TFIM case, i.e., it performs best only in a region with small $\alpha$ and large $T$. Furthermore, Magnus-THRIFT formulas of order $p>1$ would involve unitaries acting on more than 2 qubits, resulting in a higher 2-qubit gate cost.

\subsection{1D \FH model with weak hopping} \label{sec:fh}

The last model simulated for numerical tests is the \FH model on a 1D chain,
which provides an example of a fermionic simulation.
The Hamiltonian of the \FH model is
\begin{equation}
  \HFH = - \hop \sum_{\braket{i,j},\sigma} \left(c_{i,\sigma}^\dagger c_{j,\sigma} + c_{j,\sigma}^\dagger c_{i,\sigma}\right) 
           + U \sum_{i} n_{i\uparrow} n_{i \downarrow},
  \label{eq:fh_hamiltonian}
\end{equation}
where $c_{i,\sigma}^{(\dagger)}$ are the fermionic annihilation (creation) operators
on site $i$ with spin $\sigma$ and $n_{i,\sigma} = c_{i\sigma}^{\dagger}c_{i\sigma}$
are the corresponding number operators. The first sum runs over all edges $\braket{i,j}$
of the lattice and the second over all sites $i$.

In numerical simulations, 
we fix the interaction strength $U=1$, let $\alpha := -\hop$ be the small
parameter, and measure time $T$ in units of $U^{-1}$. To map the fermionic Hamiltonian to qubits, we use the Jordan-Wigner
transformation and the same circuits developed in~\cite{Cade2019}. This results 
in a ladder-like interaction graph of the qubit Hamiltonian with one rung 
corresponding to the spin-up state on a site and one to the spin-down state on
that site. As discussed in \cref{app:subsubsec:1D_FermiHubbard}, the interaction Hamiltonian $H_1(t)$ then consists of terms acting on four 
qubits, corresponding to the spin-up and spin-down states on neighbouring sites (see \cref{fig:FH_THR}).
We find numerically that time evolution with such a four-local term can 
be decomposed into a product of three evolutions with the hopping and three
evolutions with the interaction terms for all values of $T/N$ (the Trotter time step), $\hop$, and $U$. 
This means that one step of any \THR circuit takes three times the 2-qubit gate
depth of the corresponding Trotter circuit. This is in contrast to the \TFIM or Heisenberg model where the depth of \THR and Trotter methods is the same, because $H_0$ is 1-local and hence $H_1(t)$ has the same locality as $H_1$. The 2-qubit gate depths per step and number of steps used in
\cref{fig:fh_tds_algo_comparison_1x5} are shown in \cref{tab:fh_step_depths}.

\begin{figure}
\centering
\begin{minipage}[c]{.49\textwidth}
  \centering
  \input{figures/fh_tds_algo_comparison_1x5.pgf}
  \captionof{figure}{
    Landscape of the best TDS algorithm, as measured by the worst-case error
    $\Vert U - U_{\mathrm{exact}}\Vert$, as a function of the hopping strength $\hop$
    and evolution time $T$ at identical circuit depth for a $1 \times 5$ \FH chain.
    The circuit depth is
    fixed to 1 step of \THR 4 evolution. For the other algorithms, the
    number of steps is chosen to match the 2-qubit depth as closely as possible
    according to the 2-qubit depths shown in \cref{tab:fh_step_depths}.
    The colour of each point represents the algorithm that achieves the lowest
    error at those values of $\hop$ and $T$, while the brightness indicates
    the magnitude of the error.
  }
  \label{fig:fh_tds_algo_comparison_1x5}
\end{minipage}%
\hfill
\begin{minipage}[c]{.49\textwidth}
  \centering
  \input{figures/fh_thrift_depth_requirements_t=0.0625_averagecase.pgf}
  \captionof{figure}{
    2-qubit depth to achieve average infidelity
    $\mathbb{E}_{\{\ket{x}\}}[1 - |\braket{x|U_{\mathrm{exact}}^\dagger U|x}|^2] \leq  0.01$
    for the different TDS algorithms for a $1\times L$
    Fermi-Hubbard chain with an interaction strength of $\hop=1/16$ and evolution
    time $T = 2L$. Unlike \cref{fig:tfim_thrift_depth_requirements_worstcase},
    we use average fidelity to be able to simulate larger system sizes. Error bars are $\pm 1$ step and the
    shaded regions are the one-sigma confidence intervals of the extrapolations. Note that, unlike the \TFIM and Heisenberg models, the extrapolation from small sizes is not as conclusive in this case, making it difficult to determine the best-performing algorithm in the 100-qubit regime.
  }
  \label{fig:fh_thrift_depth_requirements_t=0.0625_averagecase}
\end{minipage}
\end{figure}

\begin{table}[]
  \centering
      \begin{tabular}{l|rrr}
    Algorithm           & 2-qubit gate depth & CNOT depth  & \# steps in      \cref{fig:fh_tds_algo_comparison_1x5} \\ \hline
    Trotter 1           & $3N$               & $6N$        & 20       \\
    Trotter 2           & $4N+1$             & $8N+2$      & 15       \\
    Trotter 4           & $20N+1$            & $40N+2$     & 3        \\
    optimised Trotter 8 & $60N+1$            & $120N+2$    & 1        \\
    \THR 1              & $7N$               & $14N$       & 8       \\
    \THR 2              & $8N+3$             & $16N+6$     & 7       \\
    \THR 4              & $40N+3$            & $80N+6$     & 1        \\
    optimised \THR 8    & $120N+3$           & $240N+6$    & N/A  \\
    optimised small $A$ 4 & $16N+1$            & $32N + 2$     & 3    
    \end{tabular}
    \caption{
      Circuit depth comparison of the different TDS algorithms investigated and shown in \cref{fig:fh_tds_algo_comparison_1x5} for the 1D Fermi-Hubbard model. The first column shows the 2-qubit depth of the circuit corresponding to $N$ Trotter steps in terms of arbitrary 2-qubit gates. The second column shows the corresponding cost in terms of CNOT gates. Finally, the third column gives the number of Trotter steps used in \cref{fig:fh_tds_algo_comparison_1x5}, which correspond to a fixed budget of arbitrary 2-qubit gates of 61. Note that, in the latter, we do not include the optimised \THR 8 algorithm since a single step requires deeper circuits than we allowed for \cref{fig:fh_tds_algo_comparison_1x5} and increasing the circuit depth would result in most of \cref{fig:fh_tds_algo_comparison_1x5} being dominated by the numerical noise floor.
    }
  \label{tab:fh_step_depths}
\end{table}

In \cref{fig:fh_tds_algo_comparison_1x5,fig:fh_thrift_depth_requirements_t=0.0625_averagecase}
we repeat, for 1D \FH chains, the same numerical analysis that we did for the \TFIM in 
\cref{fig:tfim_tds_algo_comparison,fig:tfim_thrift_depth_requirements_worstcase} 
and for the 1D Heisenberg model in 
\cref{fig:heisenberg_tds_algo_comparison,fig:heisenberg_thrift_depth_requirements_averagecase}. Because the \FH model
needs two qubits per site---one for each spin direction---and is not integrable, 
we are limited to much smaller system sizes, and for the depth scaling
shown in \cref{fig:fh_thrift_depth_requirements_t=0.0625_averagecase}, we again use
the average infidelity $\mathbb{E}_{\ket{x}}[1 - |\braket{x|U_{\mathrm{exact}}^\dagger U|x}|^2]$ 
instead of the more costly worst-case error $\norm{U_{\mathrm{exact}} - U}$.
We find that within the range of $T$ and $\alpha = -\hop$ that we study,
\THR methods rarely outperform ordinary Trotter methods, and in the regions they do (i.e., for $\alpha \leq 10^{-2}$)
they are beaten by the ``small $A$'' method of \citeauthor{Omelyan_2002_optimized} In particular, the 
optimised eighth-order Trotter formula of \cite{morales2022greatly} and the
``small $A$'' method of \citeauthor{Omelyan_2002_optimized} perform best out of all tested formulas for a wide range
of $T$ and $\alpha$. The main reason for the poor performance of \THR can be traced back to the high cost of implementing the gates arising in the \THR decomposition, as can be seen in \cref{tab:fh_step_depths}. Indeed, as explained in detail in \cref{app:subsubsec:1D_FermiHubbard}, the latter contains 4-local terms that are each implemented with 3 layers of arbitrary 2-qubit gates.
Similar conclusions can be drawn by looking at the 2-qubit gate depths required to achieve a fixed
average infidelity as a function of system size $L$ and evolution time $T$, as
shown in \cref{fig:fh_thrift_depth_requirements_t=0.0625_averagecase}. Even for 
$\alpha = 1/16$, the Trotter methods have lower circuit depths than 
the corresponding \THR methods. The scaling exponents with $L$ and $T$
broadly agree with those expected from the theory results in 
\cref{sec:system_size_scaling_theory,sec:commutator_scaling} and are analysed
in more detail and as a function of $\alpha$ in \cref{sec:additional_numerics_fh}.

Given the data shown in \cref{fig:fh_tds_algo_comparison_1x5}, we chose not to 
numerically study the performance of the Magnus-\THR algorithms for the \FH model.
Since \THR methods only become advantageous for $\alpha \leq 10^{-2}$ with respect to Trotter methods due to 
the more complex gates needed for the \THR circuits, and the second-order 
Magnus-\THR Hamiltonian $\Omega^{(2)}$ has up to 6-local terms 
that must be split into at least three simultaneously implementable terms
(assuming the ability to implement arbitrary 6-qubit gates), we expect that 
the values of $\alpha$ for which Magnus-\THR becomes advantageous are rather small.

\section{Discussion}

Better algorithms to simulate the time dynamics of Hamiltonians with different scales have natural applications in systems where the interactions have distinct origins.
We have shown both theoretically and through numerical experiments in various systems that the \THR algorithms can achieve better scaling than standard product formulas for Hamiltonians with different energy scales. Concretely, we consider Hamiltonians of the form $H= H_0+\alpha H_1$, where $\alpha\ll 1$ and the norms of $H_0$ and $H_1$ are comparable. Using product formulas with a carefully chosen partition, we can achieve an $O(\alpha^2t^k)$ error scaling for any $k\in \mathbb{N}$, which is better by a factor of $\alpha$ compared to the standard product formulas that do not use any structure of the Hamiltonian. We also present two algorithms to achieve scaling $O(\alpha^k t^k)$ of the approximation error. These two algorithms perform better than other formulas only in small, extreme regions of the parameter space of the systems we consider. However, such a scaling with $\alpha$ cannot be achieved using products of time-ordered evolutions according to the terms of the Hamiltonian, and they may achieve better performance in other applications.

While we have concentrated on the evolution generated by time-independent Hamiltonians, the methods developed in this work also generalise to time-dependent Hamiltonians satisfying the same assumptions. Consider a Hamiltonian $H(t)=H_0(t)+\alpha H_1(t)$, where $H_0(t)$ and $H_1(t)$ are time dependent and have similar norms for all times $t$. As before we consider $\alpha$ small. Using the same ideas developed in \cref{sec:motivation}, it is possible to show that for a partition of $H_1(t)=H_1^A(t)+H_1^B(t)$, evolving the system with the approximant
\begin{align}
U_{\rm apx}(t,0):=\mathcal{T}e^{-i\int_0^t H_0(s)ds}\mathcal{T}e^{-i\int_0^t \tilde{H}_1^A(s)ds}\mathcal{T}e^{-i\int_0^t \tilde{H}_1^B(s)ds},
\end{align}
induces an error bounded by 
\begin{align}
\| \mathcal{T}e^{-i\int_0^t H(s)ds}-U_{\rm apx}(t,0)\|\leq \alpha^2
\int_{0}^{t}dv\int_{0}^{v}ds\|[\tilde{H}_{1}^{A}(s),\tilde{H}_{1}^{B}(v)]\|,
\end{align}
where $\tilde{H}_{1}^{A,B}(t):=\mathcal{T}e^{i\int_0^t H_0(s)ds}{H}_{1}^{A,B}(t)\mathcal{T}e^{-i\int_0^t H(s)ds}$.
The main difference with respect to the time-independent case is that the evolution over a total time $T$ cannot generically be obtained from repeating the evolution over small times, but instead must be obtained from an approximation of each time-ordered slice of the total evolution.

Although these algorithms lack the competitive scaling of other approaches not based on product formulas, it has been shown \cite{Childs2018} that in the regime of medium sizes and time evolution scaling with the system size, standard product formulas can outperform asymptotically better algorithms. This makes our approach competitive in practical applications.

Developing algorithms that utilise the structure of the Hamiltonian to lower the cost of simulating time dynamics is crucial to make quantum computers useful sooner.
In particular, our approach may help to study {\it dynamical phase transitions} \cite{Heyl_2018}, where the behaviour of the dynamics of a system can change as a function of the parameters of the Hamiltonian. Quantum algorithms for time dynamics that fare well in particular regions of the parameter space allow exploring these questions with fewer resources, or for longer times given fixed resources and error.

\section*{Acknowledgements}

We thank J.\ Ostmeyer for pointing out Ref.~\cite{Omelyan_2002_optimized}. This work received funding from the European Research Council (ERC) under the European Union's Horizon 2020 research and innovation programme (grant agreement No.\ 817581), and from EPSRC grant EP/S516090/1, InnovateUK grant 44167, and InnovateUK grant 10032332. Andrew Childs's contribution to this publication was not part of his University of Maryland duties or responsibilities. 

\newpage
\appendix

\LARGE
\begin{center}
    Appendices
\end{center}
\normalsize

\section{Error scaling of \THR}

\subsection{Commutator scaling}
\label{sec:commutator_scaling}

\THR methods
approximate an interaction picture evolution unitary to $p$th order in $t$ 
via a time-dependent product formula of the form
\begin{equation}
    S_p(t)=e^{H_0t}\prod_{v=1}^\Upsilon\prod_{\gamma=1}^\Gamma \T e^{\int_{a_{v-1}t}^{a_vt} H_{\pi_v(\gamma)}(s)ds}.
\end{equation}

Note that, to reduce clutter and to avoid keeping track of phases, the factors of $i$ are absorbed into the Hamiltonians in the following analysis. The results are unaffected by this choice.

Using the fact that the time dependence of the $H_\gamma(t)$ is simply unitary evolution under $H_0$,
this is converted back to an equivalent product formula of time-independent terms
\begin{equation}
    S_p(t) =\prod_{v=1}^\Upsilon e^{(H_{\pi_v(\Gamma)}+H_0)A_vt}e^{-H_0A_vt}e^{(H_{\pi_v(\Gamma-1)}+H_0)A_vt}
    e^{-H_0A_vt}\cdots e^{-H_0A_vt}e^{(H_{\pi_v(1)}+H_0)A_vt},
\end{equation}
where $A_v=a_v-a_{v-1}$ with $a_0=0$. This is essentially a Trotter-style product 
formula of the Hamiltonian $H=H_0+\sum_{\gamma=1}^\Gamma H_\gamma$
where $H$ is decomposed into the sum 
\begin{equation}
    H=\sum_{l=1}^{2\Gamma-1}\tilde{H}_{l} 
\end{equation}
with $\tilde{H}_{2\gamma-1}=H_\gamma+H_0$ and $\tilde{H}_{2\gamma}=-H_0$.
This product formula fits the general form used in \cite{Childs2021},
\begin{equation}
    S_p(t)=\prod_{v=1}^\Upsilon\prod_{l=1}^{2\Gamma-1} e^{\tilde{H}_{\pi_v(l)}A_vt},
\end{equation}
and by the main result of that work, the additive and multiplicative errors $\mathcal{A}(t)$, 
$\mathcal{M}(t)$, defined as
\begin{equation}
    S_p(t) = e^{Ht}+\mathcal{A}(t) = e^{Ht}(I+\mathcal{M}(t)),
\end{equation}
both scale as
\begin{equation}\label{eq: THRIFT scaling big commutator}
    \|\mathcal{A}(t)\|,\|\mathcal{M}(t)\|=O(\tilde{T}_pt^{p+1}),
\end{equation}
where
\begin{equation}
    \tilde{T}_p=\sum_{l_1,\dots,l_{p+1}=1}^{2\Gamma-1}\|
    [\tilde{H}_{l_{p+1}},\cdots[\tilde{H}_{l_2},\tilde{H}_{l_1}]\cdots]\|,
\end{equation}
with $\|\cdot\|$ denoting spectral norm. Expanding commutators containing terms of the form $H_\gamma+H_0$ and applying the triangle 
inequality, we have
\begin{equation}
    \tilde{T}_p \leq \sum_{\gamma_1,\dots,\gamma_{p+1}=0}^{\Gamma}C_{\gamma_1,\dots,\gamma_{p+1}}\|
    [{H}_{\gamma_{p+1}},\cdots[{H}_{\gamma_2},{H}_{\gamma_1}]\cdots]\|,
\end{equation}
where the sum is now over $\{H_0,\dots,H_\Gamma\}$ and $C_{\gamma_1,\dots,\gamma_{p+1}}$ are constants.
Setting $C_{\Gamma,p}=\max\{C_{\gamma_1,\dots,\gamma_{p+1}}\}$ (dependent only on $\Gamma$ and $p$) and defining
\begin{equation}
    T_p = \sum_{\gamma_1,\dots,\gamma_{p+1}=0}^{\Gamma}\|
    [{H}_{\gamma_{p+1}},\cdots[{H}_{\gamma_2},{H}_{\gamma_1}]\cdots]\|,
\end{equation}
we have
\begin{equation}
    \tilde{T}_p\leq C_{\Gamma,p}T_p.
\end{equation}
Taking $\Gamma$ and $p$ as constant, we then have for a $p$th-order \THR
product formula
\begin{equation}\label{eq: THRIFT scaling small commutator}
    \|\mathcal{A}(t)\|,\|\mathcal{M}(t)\|=O({T}_pt^{p+1}),
\end{equation}
i.e., the same asymptotic scaling as a standard product formula for the decomposition
into $\{H_0,\dots,H_\Gamma\}$.

A commutator treatment for average-case product formula error is given in \cite{zhao2022hamiltonian}. They show that for a $p$th-order product formula approximating evolution under $\sum_{\gamma=0}^\Gamma H_\gamma$ applied to states drawn from a 1-design input ensemble, the average error in the $l_2$ norm is bounded asymptotically as
\begin{equation}
    R_{l_2}=O(T^F_pt^{p+1}),
\end{equation} 
where
\begin{equation}
    T^F_p=\sum_{\gamma_1,\dots,\gamma_{p+1}=0}^{\Gamma}\frac{1}{\sqrt{d}}\|
    [{H}_{\gamma_{p+1}},\cdots[{H}_{\gamma_2},{H}_{\gamma_1}]\cdots]\|_F,
\end{equation}
with $\|H\|_F=\sqrt{\text{Tr}[HH^\dagger]}\leq \sqrt{d}\|H\|$ denoting the Frobenius norm. By the same argument as above for the spectral error, this asymptotic bound applies equally to THRIFT.

\subsection{System size scaling for geometrically local Hamiltonians}
\label{sec:system_size_scaling_theory}

Given a $d$-dimensional lattice $\Lambda^d$ of $n$ qubits with distance metric $D$, define a
geometrically local Hamiltonian as
\begin{equation}\label{eq: lattice H}
    H=\sum_{Z\subset \Lambda^d}H_Z
\end{equation}
where $H_Z$ acts only on a finite subset of lattice sites $Z$ and there exists
a constant, finite $R$ such that
    \begin{equation}
        \|H_Z\|\leq \begin{cases}
            1 & \text{if } \diam(Z)\leq R \\
            0 & \text{if } \diam(Z)>R,
        \end{cases}
    \end{equation}
where $\diam(Z)=\max\{D(i,j):i,j\in Z\}$ is the maximum distance between
any two points in $Z$.

\begin{lemma}
    A $p$th-order product formula approximating evolution under a Hamiltonian  $H=\sum_{\gamma=1}^\Gamma H_\gamma$,
    where all $H_\gamma$ are geometrically local on a lattice of $n$ qubits, has additive and multiplicative
    error with the following asymptotic scaling:
\begin{equation}
    \|\mathcal{A}(t)\|,\|\mathcal{M}(t)\|=O(n t^{p+1}).
\end{equation}
\end{lemma}

\begin{proof}
     By the results in
    \cite{Childs2021} we have the bound
    \begin{equation}\label{eq:Add Mult Error}
        \|\mathcal{A}(t)\|,\|\mathcal{M}(t)\|=O\left(
        \sum_{\gamma_1,\dots,\gamma_{p+1}=1}^\Gamma
        \|W_{\gamma_1,\dots,\gamma_{p+1}}\|t^{p+1}
        \right),
    \end{equation}
    where $W_{\gamma_1,\dots,\gamma_{p+1}}=[H_{\gamma_{p+1}},\dots[H_{\gamma_2},H_{\gamma_1}]\dots]$. 
    As any given $H_\gamma$ is geometrically local it can be written as in \cref{eq: lattice H} as
    \begin{equation}
    H=\sum_{Z\subset \Lambda^d}H_{\gamma,Z},
    \end{equation}
    where the $H_{\gamma,Z}$ act on subsets $Z$ of maximum diameter $R$. For lattice site $i$ let us define
    \begin{equation}
        H_\gamma^i:=\sum_{Z\ni i}\frac{1}{|Z|}H_{\gamma,Z},
    \end{equation}
    i.e., the sum of all local terms in $H_\gamma$ that act on site $i$, each divided by the size of their support set; this accounts for multi-counting and means that we can write
    \begin{equation}
        H_\gamma = \sum_{i\in\Lambda^d}H_\gamma^i.
    \end{equation}
    For each $H_\gamma^i$ we have $\|H^i_\gamma\|\leq C$ 
    for some constant $C$ dependent on $R$ and $d$. We may now write
    \begin{equation}
        W_{\gamma_1,\dots,\gamma_{p+1}}=\sum_{i_1,\dots,i_{p+1}\in\Lambda^d}
        [H^{i_{p+1}}_{\gamma_{p+1}},\dots[H^{i_2}_{\gamma_2},H^{i_1}_{\gamma_1}]\dots].
    \end{equation}
    We can simplify this expression by omitting terms that are zero due to lack of
    shared support. The commutator $[H^{i_2}_{\gamma_2},H^{i_1}_{\gamma_1}]$ vanishes
    if $i_2$ is more than $2R$ away from $i_1$ as no part of the two arguments will
    overlap. Furthermore, assuming the inside commutator is nonzero, 
    $[H_{\gamma_3}^{i_3},[H^{i_2}_{\gamma_2},H^{i_1}_{\gamma_1}]]$
    vanishes if $i_3$ is more than $3R$ away from $i_1$, because at that distance,
    $H_{\gamma_3}^{i_3}$ only overlaps with the parts of $H^{i_2}_{\gamma_2}$
    that do not overlap with $H_{\gamma_1}^{i_1}$. By similar logic, $i_4$ must be
    within $4R$ of $i_1$, and so on. We can then reduce the sum to
    \begin{equation}
        W_{\gamma_1,\dots,\gamma_{p+1}}=\sum_{i_{p+1}:D(i_{p+1},i_1)\leq(p+1)R}\cdots
        \sum_{i_2:D(i_2,i_1)\leq2R}\sum_{i_1\in\Lambda^d}
        [H^{i_{p+1}}_{\gamma_{p+1}},\dots,[H^{i_2}_{\gamma_2},H^{i_1}_{\gamma_1}]\dots].
    \end{equation}
    The number of lattice points within a fixed distance of a given point is constant,
    so the sum over $\|W_{\gamma_1,\dots,\gamma_{p+1}}\|$ in \cref{eq:Add Mult Error} simply reduces to a sum of constants over the points $i_1\in\Lambda^d$,
    which is proportional to the number of lattice points $n$. The result follows.
\end{proof}

\begin{corollary}
    For a $p$th-order product formula for a Hamiltonian $H$ as described above, to
    simulate evolution for time $t$ with accuracy $\epsilon$, it suffices to use $r$ iterations of 
    the product formula, where
    \begin{equation}
        r=O\left(\frac{n^{1/p}}{\epsilon^{1/p}}t^{1+1/p}\right).
    \end{equation}
\end{corollary}
The above results apply equally to THRIFT product formulas, using either the error scaling 
given in \cref{eq: THRIFT scaling big commutator} or in \cref{eq: THRIFT scaling small commutator}. Furthermore, linear scaling of the spectral error with system size also applies to the average-case error by a similar argument.

\subsection{Limits on error scaling in \texorpdfstring{$\alpha$}{alpha} for time-dependent product formulas}\label{appendix_2}

In this section we establish limitations on how well time-dependent product formulas can approximate Hamiltonian dynamics as a function of $\alpha$, a scaling factor for the Hamiltonian. Such an evolution is obtained in \THR when approximating the time-dependent part of an interaction-picture evolution operator via a time-dependent product formula. In particular, the Hamiltonian in this time-dependent part is scaled by $\alpha$, so these results provide limitations on the $\alpha$-dependence of \THR.

\begin{theorem}\label{thm: alpha squared scaling general}
    For a Hamiltonian of the form
\begin{equation}
    H(t) = \sum_{\gamma=1}^\Gamma \alpha H_\gamma(t),
\end{equation}
consider a time-dependent product formula of the form
\begin{equation}\label{eq:prod formula}
	S(t)=\prod_{v=1}^\Upsilon\prod_{\gamma=1}^\Gamma \T e^{\int_{a_{v-1}t}^{a_vt}\alpha H_{\pi_v(\gamma)}(s)ds},
\end{equation}
where $\pi_v$ are permutations of the indices $\gamma$ and $a_v$ are real numbers defining time intervals $[a_{v-1}t,a_v t]$. 
There is no such product formula for which
\begin{equation}\label{eq: product formula error}
		\left\|S(t)-\T e^{\int_{0}^{t}\alpha H(s)ds}\right\|=O(\alpha^k)
	\end{equation}
for $k>2$ and $t\neq 0$.
\end{theorem}

\begin{proof}
We expand the terms of \cref{eq: product formula error} into integral series, yielding Taylor series in $\alpha$ that can be compared term-by-term. Up to second order, the Dyson series for evolution under $H(t)$ over the interval $[0,t]$ is
\begin{equation}
    \T e^{\int_{0}^{t}\alpha H(s)ds} = 1 + \alpha \int_0^tds_1\sum_{\gamma} H_\gamma(s_1)+\alpha^2\int_0^tds_1\int_0^{s_1}ds_2\sum_{\gamma_1,\gamma_2} H_{\gamma_1}(s_1)H_{\gamma_2}(s_2)+O(\alpha^3).
\end{equation}
Expanding each time-ordered integral in $S(t)$ and collecting powers of $\alpha$ gives the second-order expansion
\begin{equation}
\begin{split}
    S(t) &= 1 + \alpha \sum_{v} \int_{a_{v-1}t}^{a_vt}ds_1\sum_{\gamma}H_\gamma(s_1) \\
    &\quad+ \alpha^2 \sum_{v} \int_{a_{v-1}t}^{a_vt}ds_1\int_{a_{v-1}t}^{s_1}ds_2\sum_\gamma H_\gamma(s_1)H_\gamma (s_2) \\ 
    &\quad+ \alpha^2\sum_v\int_{a_{v-1}t}^{a_vt}ds_1\int_{a_{v-1}t}^{a_vt}ds_2\sum_{\gamma_1<\gamma_2}H_{\pi_v(\gamma_1)}(s_1)H_{\pi_v(\gamma_2)}(s_2)  \\
    &\quad+\alpha^2 \sum_{v>u}\int_{a_{v-1}t}^{a_vt}ds_1\int_{a_{u-1}t}^{a_ut}ds_2\sum_{\gamma_1,\gamma_2}H_{\gamma_1}(s_1)H_{\gamma_2}(s_2)+O(\alpha^3).
\end{split}
\end{equation}
Combining integrals with matching boundaries and using the fact that $v>u$, we may rewrite this as
\begin{equation}
\begin{split}
    S(t) &= 1 + \alpha  \int_{a_{0}t}^{a_\Upsilon t}ds_1\sum_{\gamma}H_\gamma(s_1) \\
    &\quad+ \alpha^2  \int_{a_{0}t}^{a_\Upsilon t}ds_1\int_{a_{0}t}^{s_1}ds_2\sum_\gamma H_\gamma(s_1)H_\gamma (s_2) \\ 
    &\quad+ \alpha^2\sum_v\int_{a_{v-1}t}^{a_vt}ds_1\int_{a_{v-1}t}^{a_vt}ds_2\sum_{\gamma_1<\gamma_2}H_{\pi_v(\gamma_1)}(s_1)H_{\pi_v(\gamma_2)}(s_2)  \\
    &\quad+\alpha^2 \sum_{v}\int_{a_{0}t}^{a_vt}ds_1\int_{a_{0}t}^{a_{v-1}t}ds_2\sum_{\gamma_1\neq \gamma_2}H_{\gamma_1}(s_1)H_{\gamma_2}(s_2)+O(\alpha^3).
\end{split}
\end{equation}
Clearly, for $S(t)$ to approximate evolution under $H(t)$ over $[0,t]$ to first order in $\alpha$, we must have $a_0=0$ and $a_\Upsilon=1$. Then we find the following expression for the difference between $S(t)$ and the ideal evolution:
\begin{equation}
\begin{split}
    S(t)-\T e^{\int_{0}^{t}\alpha H(s)ds}&=\alpha^2\sum_v\int_{a_{v-1}t}^{a_vt}ds_1\int_{a_{v-1}t}^{a_vt}ds_2\sum_{\gamma_1<\gamma_2}H_{\pi_v(\gamma_1)}(s_1)H_{\pi_v(\gamma_2)}(s_2)   \\ 
    &\quad+\alpha^2 \sum_{v}\int_{0}^{a_vt}ds_1\int_{0}^{a_{v-1}t}ds_2\sum_{\gamma_1\neq \gamma_2}H_{\gamma_1}(s_1)H_{\gamma_2}(s_2) \\
    &\quad-\alpha^2\int_0^tds_1\int_0^{s_1}ds_2\sum_{\gamma_1\neq \gamma_2} H_{\gamma_1}(s_1)H_{\gamma_2}(s_2)+O(\alpha^3).
\end{split}
\end{equation}
By inserting 
\begin{equation}
0=\alpha^2\sum_v\int_{a_{v-1}t}^{a_vt}ds_1\int_{a_{v-1}t}^{s_1}ds_2\sum_{\gamma_1<\gamma_2}\left(H_{\pi_v(\gamma_2)}(s_1)H_{\pi_v(\gamma_1)}(s_2)- H_{\pi_v(\gamma_2)}(s_1)H_{\pi_v(\gamma_1)},(s_2)\right),
\end{equation}
doing some algebra, and relabelling integral variables where needed, we finally arrive at a more compact form for the error at second order, namely
\begin{equation}
\begin{gathered}
    S(t)-\T e^{\int_{0}^{t}\alpha H(s)ds}=\Delta\alpha^2 + O(\alpha^3) \\
    \text{with}\quad\Delta= \sum_v \int_{a_{v-1}t}^{a_vt}ds_1\int_{s_1}^{a_vt}ds_2\sum_{\gamma_1<\gamma_2}[H_{\pi_v(\gamma_1)}(s_1),H_{\pi_v(\gamma_2)}(s_2)].
\end{gathered}
\end{equation}
The question then becomes: is there a generic set of parameters $\{a_v\}$ and $\{\pi_v\}$ such that $\Delta$ vanishes?\footnote{Technically, for the error to agree up to second order, $\Delta$ only needs to be $O(\alpha^3)$, but for a generic set of parameters there can be no $\alpha$ dependence.}
\begin{table}
	\centering
	\begin{tabular}{ccc}
		\centering
		category & function definition & non-zero region \\
		\hline
		\makecell{$\Pi^v_{\gamma_1,\gamma_2}$ even\\ $a_{v-1}t<a_vt$} & $f^v_{\gamma_1,\gamma_2}(s_1,s_2)=\begin{cases}
			1\;\;:&\;\;{a_{v-1}t<s_1<s_2<a_vt} \\
			0\;\;:&\;\;\mbox{otherwise}
		\end{cases}$ & \makecell{\begin{tikzpicture}
				\draw (0,0) rectangle (1,1);
				\fill[blue!40!white] (0,0)--(1,1)--(0,1)--cycle;
				\draw[thick,->,line cap=round] (0,0)--(1.2,0) node[anchor=west]{\small$s_1$};
				\draw[thick,->,line cap=round] (0,0)--(0,1.2) node[anchor=south]{\small$s_2$};
				\node at (.3,.7){\textbf{+}};
				\node at (-.2,-.2){\tiny$a_{v-1}t$};
				\node[rotate=90] at (-.2,1){\tiny$a_vt$};
				\node at (1,-.2){\tiny$a_vt$};
		\end{tikzpicture}}  \\
		\makecell{$\Pi^v_{\gamma_1,\gamma_2}$ even\\ $a_vt<a_{v-1}t$} & $f^v_{\gamma_1,\gamma_2}(s_1,s_2)=\begin{cases}
			1\;\;:&\;\;{a_vt<s_2<s_1<a_{v-1}t} \\
			0\;\;:&\;\;\mbox{otherwise}
		\end{cases}$ & \makecell{\begin{tikzpicture}
				\draw (0,0) rectangle (1,1);
				\fill[blue!40!white] (0,0)--(1,1)--(1,0)--cycle;
				\draw[thick,->,line cap=round] (0,0)--(1.2,0) node[anchor=west]{\small$s_1$};
				\draw[thick,->,line cap=round] (0,0)--(0,1.2) node[anchor=south]{\small$s_2$};
				\node at (.7,.3){\textbf{+}};
				\node at (-.2,-.2){\tiny$a_vt$};
				\node[rotate=90] at (-.2,1){\tiny$a_{v-1}t$};
				\node at (1,-.2){\tiny$a_{v-1}t$};
		\end{tikzpicture}}  \\
		\makecell{$\Pi^v_{\gamma_1,\gamma_2}$ odd\\ $a_{v-1}t<a_vt$} & $f^v_{\gamma_1,\gamma_2}(s_1,s_2)=\begin{cases}
			-1\;\;&:\;\;{a_{v-1}t<s_2<s_1<a_vt} \\
			\hphantom{-}0\;\;&:\;\;\mbox{otherwise}
		\end{cases}$ & \makecell{\begin{tikzpicture}
				\draw (0,0) rectangle (1,1);
				\fill[red!40!white] (0,0)--(1,1)--(1,0)--cycle;
				\draw[thick,->,line cap=round] (0,0)--(1.2,0) node[anchor=west]{\small$s_1$};
				\draw[thick,->,line cap=round] (0,0)--(0,1.2) node[anchor=south]{\small$s_2$};
				\node at (.7,.3){\textbf{--}};
				\node at (-.2,-.2){\tiny$a_{v-1}t$};
				\node[rotate=90] at (-.2,1){\tiny$a_vt$};
				\node at (1,-.2){\tiny$a_vt$};
		\end{tikzpicture}} \\
		\makecell{$\Pi^v_{\gamma_1,\gamma_2}$ odd\\ $a_vt<a_{v-1}t$} & $f^v_{\gamma_1,\gamma_2}(s_1,s_2)=\begin{cases}
			-1\;\;&:\;\;{a_vt<s_1<s_2<a_{v-1}t} \\
			\hphantom{-}0\;\;&:\;\;\mbox{otherwise}
		\end{cases}$ & \makecell{\begin{tikzpicture}
				\draw (0,0) rectangle (1,1);
				\fill[red!40!white] (0,0)--(1,1)--(0,1)--cycle;
				\draw[thick,->,line cap=round] (0,0)--(1.2,0) node[anchor=west]{\small$s_1$};
				\draw[thick,->,line cap=round] (0,0)--(0,1.2) node[anchor=south]{\small$s_2$};
				\node at (.3,.7){\textbf{--}};
				\node at (-.2,-.2){\tiny$a_vt$};
				\node[rotate=90] at (-.2,1){\tiny$a_{v-1}t$};
				\node at (1,-.2){\tiny$a_{v-1}t$};
		\end{tikzpicture}} \\
	\end{tabular}
	\caption{The form of $f^v_{\gamma_1,\gamma_2}$ depends on the parity of $\Pi^v_{\gamma_1,\gamma_2}$, i.e., whether it switches 1 and 2, and whether $a_{v-1}t<a_vt$. All four possibilities are shown, along with a visualisation of their non-zero region as a shaded area on the $s_1$-$s_2$ plane. Blue shading with a ``$+$'' indicates a value of $+1$ and, likewise, red with a ``$-$'' a value of $-1$.}
	\label{tab: ints to triangles}
\end{table}
Relabelling variables and using the antisymmetry of the commutator, we may write
\begin{equation}\label{eq: DELTA same function}
	\Delta = \sum_{v }\sum_{\gamma_1<\gamma_2}\mathcal{S}(\pi_v,\gamma_1,\gamma_2)\int_{a_{v-1}t}^{a_vt}ds_{\Pi^v_{\gamma_1,\gamma_2}(1)}\int_{s_{\Pi^v_{\gamma_1,\gamma_2}(1)}}^{a_vt}ds_{\Pi^v_{\gamma_1,\gamma_2}(2)}[H_{\gamma_1}(s_1),H_{\gamma_2}(s_2)]
\end{equation}
where the function $\mathcal{S}(\pi_v,\gamma_1,\gamma_2)$ is $-1$ if the permutation $\pi_v$ switches the order of $\gamma_1$, $\gamma_2$ and $+1$ otherwise, and $\Pi^v_{\gamma_1,\gamma_2}$ is a permutation of the indices 1 and 2 defined as
\begin{equation}
    \Pi^v_{\gamma_1,\gamma_2}((1,2))=\begin{cases}
        (1,2)& \text{if }\pi_v((\gamma_1,\gamma_2))=(\gamma_1,\gamma_2),\\
        (2,1)&\text{if }\pi_v((\gamma_1,\gamma_2))=(\gamma_2,\gamma_1).
    \end{cases}
\end{equation}
We may further rewrite this as
\begin{equation}\label{eq: DELTA weighted integral}
	\Delta = \sum_{\gamma_1<\gamma_2}\;\;\iint\limits_{[t_{\text{min}},t_{\text{max}}]^2}ds_1ds_2\sum_{v} f^v_{\gamma_1,\gamma_2}(s_1,s_2)[H_{\gamma_1}(s_1),H_{\gamma_2}(s_2)],
\end{equation}
where $t_\text{min}$ and $t_\text{max}$ are the minimum and maximum values of $\{a_vt\}_{v=1}^\Upsilon$ and $f^v_{\gamma_1,\gamma_2}$ are functions that are zero everywhere except for the corresponding region of integration in \cref{eq: DELTA same function} where they take the value of $\mathcal{S}(\pi_v,\gamma_1,\gamma_2)$. The possible forms of these functions are visualised in \cref{tab: ints to triangles}. For $\Delta$ to vanish, each term in the sum over $\gamma_1<\gamma_2$ must vanish. Furthermore, for arbitrary time-dependent Hamiltonians, this requires that the respective sums $\sum_{v} f^v_{\gamma_1,\gamma_2}(s_1,s_2)$ vanish for all $s_1,s_2$. By inspection of the forms of these functions in \cref{tab: ints to triangles}, it is clear that this requires every time step $(a_{v-1}t,a_vt)$ to have a step of the same size in the reverse direction, meaning that $\Delta$ only vanishes for evolution over zero time. \Cref{thm: alpha squared scaling general} then follows.
\end{proof}

In fact, \cref{thm: alpha squared scaling general} holds even with the further restriction that the time-dependence of the Hamiltonian arises via conjugation by some fixed Hamiltonian dynamics, as in the case of \THR.

\begin{theorem}\label{thm: alpha squared scaling THRIFT}
   \Cref{thm: alpha squared scaling general} also holds for the restricted case where
   $H_\gamma(t) = e^{H_0 t}H_\gamma(0)e^{-H_0t}$ for some fixed $H_0$.
\end{theorem}

We show this by adapting the above proof of \cref{thm: alpha squared scaling general}.
The sum over $\gamma_1<\gamma_2$ must still vanish term-wise, so let us simplify by analysing the $\gamma_1=1$, $\gamma_2=2$ term, denoting it $T(H_1,H_2)$. Let us simplify further by writing the sum over $f^v_{1,2}$ as a single function $F$ and relabelling $s_1\rightarrow x$, $s_2\rightarrow y$, so we have
\begin{equation}
    T(H_1,H_2)=\iint\limits_{S^2}dxdyF(x,y)[H_{1}(x),H_{2}(y)],
\end{equation}
where $S:=[t_\text{min},t_\text{max}]$. We have
\begin{equation}\label{eq: interaction picture}
	H_1(t)=e^{H_0t}H_1(0)e^{-H_0t},\quad H_2(t)=e^{H_0t}H_2(0)e^{-H_0t}.
\end{equation}

\begin{lemma}
	If $T(H_1,H_2)=0$ for any $H_1$ with $[H_0,H_1(0)]\neq 0$, then $T(H'_1,H_2)=0$ for any $H'_1$ with $[H_0,H'_1(0)]=0$.
\end{lemma}
\begin{proof}
    For $H'_1$ such that $[H_0,H'_1(0)]=0$ define $H^{\prime\prime}_1(x)=H'_1(x)+A(x)$ where $A(x)$ is some operator-valued function with $[H_0,A(0)]\neq 0$. Then by linearity we have
    \begin{equation}
        T(H'_1,H_2)=T(H^{\prime\prime}_1,H_2)+T(A,H_2)=0.
    \end{equation}
    As $T(A,H_2)=0$ by hypothesis, we have $T(H'_1,H_2)=0.$
\end{proof}

\begin{lemma}\label{lem: F against monomials}
	If $T(H_1,H_2)=0$ for any $H_1,H_2$, then for all $k\in\mathbb{N}$,
	\begin{equation}
		\iint\limits_{S^2}dxdyF(x,y)x^k=\iint\limits_{S^2}dxdyF(x,y)y^k=0.
	\end{equation}
\end{lemma}
\begin{proof}
	Let $[H_1(0),H_0]=0$ and $H_0=\lambda H'_0$. Expand $\Delta$ as a power series in $\lambda$ to get
	\begin{equation}
		\Delta=\sum_k^\infty \frac{1}{k!}\iint\limits_{S^2}dxdyF(x,y)[H_1,\text{ad}_{H'_0}^kH_2]y^k\lambda^k=0
	\end{equation}
	where $\text{ad}_AB=[A,B]$ and we write $H_1(0)$ as $H_1$, likewise for $H_2$. This series must vanish term-by-term with $\lambda$, so we have
	\begin{equation}
		[H_1,\text{ad}_{H'_0}^kH_2]\iint\limits_{S^2}dxdyF(x,y)y^k=0\quad \forall k\in\mathbb{N}.
	\end{equation}
	Let $H'_0, H_1, H_2$ be Paulis that pairwise anticommute except $H_0$ and $H_1$. Then $\text{ad}_{H'_0}^kH_2=2^kH^{\prime k}_0H_2$, meaning $[H_1,\text{ad}_{H'}^kH_2]\neq 0$ for all $k\in\mathbb{N}$. For $T(H_1,H_2)=0$ to hold in general, the above integral must then vanish for all $k$. The same argument applies for $x$.
\end{proof}

\begin{lemma}\label{lem:almost everywhere}
	If for all $k\in\mathbb{N}$,
 \begin{equation}
     \int\limits_{[a,b]}dxf(x)x^k=0,
 \end{equation}
    then
	\begin{equation}
		\begin{split}
			f(x)=0\quad \text{almost everywhere in $[a,b]$},
		\end{split}
	\end{equation}
	i.e., it is nonzero on only a measure-zero subset of $[a,b]$.
\end{lemma}

\begin{proof}
    The following proof is reproduced from \cite{AEproof} in more detail.
	
	Let $f(x)$ be integrable over the interval $[a,b]$ and have the property that $\int_{[a,b]}dxf(x)x^k=0$ for all $k\in\mathbb{N}$. It follows that $\int_{[a,b]}dxf(x)p(x)=0$ for any polynomial $p$. The polynomials are dense in the set of continuous functions on $[a,b]$, so it also follows that for any continuous function $g$, $\int_{[a,b]}dxf(x)g(x)=0$.

	Assume now that $f$ is not zero almost everywhere. Then the set $\{x:f(x)>0\}$ has finite measure. One can then find $\delta$ such that $\{x:f(x)>\delta\}$ has finite measure. By the regularity of the Lebesgue measure, one can choose a compact set $K$ and open set $V$ such that $K\subset\{x:f(x)>\delta\}\subset V$ and the measure of $V\setminus K$ is arbitrarily small. Let $g$ be a continuous function such that $0\leq g(x)\leq 1$ which is equal to 1 on $K$ and 0 outside of $V$. Then we have
	\begin{equation}
		\left|\int_{[a,b]}dxg(x)f(x)\right|=\left|\int_{V}dxg(x)f(x)\right|=\left|\int_{K}dxg(x)f(x)+\int_{V\setminus K}dxg(x)f(x)\right|.
	\end{equation}
	By the (reverse) triangle inequality, we have
	\begin{equation}
		\begin{split}
			\left|\int_{[a,b]}dxg(x)f(x)\right|&\geq \left|\int_{K}dxg(x)f(x)\right|-\left|\int_{V\setminus K}dxg(x)f(x)\right|\\
		&\geq \left|\int_{K}dxg(x)f(x)\right|-\int_{V\setminus K}dx\left|g(x)f(x)\right|.
		\end{split}
	\end{equation}

 Now as $g$ is 1 on $K$ and $g<1$ for some region on $V$, we have
	\begin{equation}
		\left|\int_{[a,b]}dxg(x)f(x)\right|\geq\left|\int_{K}dxf(x)\right|-\int_{V\setminus K}dx\left|f(x)\right|.
	\end{equation}
	As noted before, $f>\delta$ on $K$, so
	\begin{equation}
		\left|\int_{[a,b]}dxg(x)f(x)\right|\geq\delta m(K)-\int_{V\setminus K}dx\left|f(x)\right|
	\end{equation}
	where $m(K)>0$ is the measure of $K$. Since $V\setminus K$ can be made arbitrarily small, we can take the right-hand side of the inequality to be positive, meaning that $\left|\int_{[a,b]}dxg(x)f(x)\right|$ for some continuous $g$, which is a contradiction.
 \end{proof}

\begin{corollary}\label{cor:F almost everywhere}
If for all $k\in\mathbb{N}$,
\begin{equation}
    \iint\limits_{S^2}dxdyF(x,y)x^k=0,
\end{equation}
    then as a function of $x$,
    \begin{equation}
        \int\limits_SdyF(x,y)=0\quad \text{almost everywhere in $S$}.
    \end{equation}
    The same holds when $y$ and $x$ are exchanged.
\end{corollary}

\begin{proof}
    Set
	\begin{equation}
		f(x)=\int_S dyF(x,y)
	\end{equation} 
    and $[a,b]=S$. The result follows immediately from \cref{lem:almost everywhere}.
\end{proof}
	
\begin{lemma}\label{lem: tUps=0}
	If $T(H_1,H_2)=0$ for all $H_0, H_1(x), H_2(y)$ as defined above, then $a_\Upsilon t=a_0t$. 
\end{lemma}
\begin{proof}
	Assume $a_\Upsilon t\neq a_0t$. The intervals $[a_{v-1}t,a_vt]$ are ``steps'' in a path from $a_0t$ to $a_{\Upsilon}t$, so any point in $S\setminus\{a_vt\}$ is contained within an odd number of these intervals as the path must cross it an odd number of times. Consider the open interval $(a_vt, a'_vt)$ where $a'_v=\min\{a_u:a_u>a_v\}$. Any point in this set is contained in the same set of ``step'' intervals. There must exist a $v$ for which $(a_vt, a'_vt)\cap[a_0t,a_\Upsilon t]\neq\varnothing$, so there exists a finite-measure set of points that are all contained within the same set of steps $M$ where $|M|$ is odd.
	
	As $F(x,y)$ is a sum of the functions in \cref{tab: ints to triangles}, we can see that for a point $x'$ in this set, the integral $\int_SdyF(x',y)$ takes the form
	\begin{equation}
		 \int_SdyF(x',y)=\sum_{v:[a_{v-1}t,a_vt]\in M}s_v(x'-T_v)
	\end{equation}
	where $s_v\in\{\pm 1\}$ and $T_v\in\{a_{v-1}t,a_vt\}$. Consider now the integral for $x'+c\in (a_vt, a'_vt)$,
	\begin{equation}
		\begin{split}
			\int_SdyF(x'+c,y)&=\sum_{v:[a_{v-1}t,a_vt]\in M}s_v(x'+c-T_v)\\
			&=\sum_{v:[a_{v-1}t,a_vt]\in M}s_v(x'-T_v)+\sum_{v:[a_{v-1}t,a_vt]\in M}s_vc.
		\end{split}
	\end{equation}
	As $|M|$ is odd and $s_v$ are signs,
	\begin{equation}
		\left|\sum_{v:[a_{v-1}t,a_vt]\in M}s_vc\right|\geq 1,
	\end{equation}
	and $\int_SdyF(x,y)$ is non-zero on a finite measure set, contradicting \cref{cor:F almost everywhere}.
\end{proof}

We are now ready to prove \cref{thm: alpha squared scaling THRIFT}.

\begin{proof}[Proof of \cref{thm: alpha squared scaling THRIFT}]
    Recall that for the product formula in \cref{eq:prod formula} to agree up to $\alpha^2$ for arbitrary times, the quantity
    \begin{equation}
        \Delta = \sum_{\gamma_1<\gamma_2}\iint\limits_{S^2}ds_1ds_2F(s_1,s_2)[H_{\gamma_1},H_{\gamma_2}]
    \end{equation}
    must vanish. This sum must vanish term-wise, so it suffices to consider the $\gamma_1=1$, $\gamma_2=2$ case. By \cref{lem: tUps=0}, this term may only vanish if $a_0t=a_\Upsilon t$. As argued in the proof of \cref{thm: alpha squared scaling general}, first-order agreement requires $a_0=0$ and $a_\Upsilon=1$, so second-order agreement can only hold if $t=0$.
\end{proof}

\section{Convergence of Magnus expansion}\label{app:convergence}

A simple proof of the convergence of the Magnus expansion is given in \cite{moan1998efficient}. Here we reproduce it for completeness. The first ingredient is the following lemma.

\begin{lemma}[A Bihari-type inequality \cite{moan1998efficient}]\label{lemma_bihari}
    Let $h, v \in C(0,T)$ (where $C(0,T)$ denotes functions with a continuous first derivative on the interval $[0,T]$) be integrable positive functions and let $g \in C(0,T)$ be a non-decreasing positive function. Then
    \begin{align}\label{eq:Bihari}
    h(x)\leq \int_0^xv(s)g(h(s))ds
    \end{align}
    for $x\in [0,T]$
    implies that $h(x)\leq \int_0^xv(s)g(h(s))ds\leq G^{-1}(\int_0^x v(s)ds)$, where $G^{-1}$ is the inverse function of $G(s)=\int_0^s\frac{ds}{g(s)}$.
\end{lemma}

\begin{proof}
Define $f(x):=\int_0^xv(x)g(h(x))dx$, so $\frac{df}{dx}=v(x)g(h(x))$. Using \cref{eq:Bihari}, $h\leq f$, which implies $g(h(x))\leq g(f(x))$ as $g$ is non-decreasing. Therefore
$\frac{df}{dx}\leq v(x)g(f(x))$. Dividing by $g$ and integrating by substitution, we have 
\begin{align}
\int_0^{f(t)}\frac{ds}{g(s)}
\leq \int_0^t v(x)dx
\quad\Rightarrow\quad 
G(f(t))\leq \int_0^tv(x)dx.
\end{align}
Applying the inverse of $G$ and using $h\leq f$ completes the proof.
\end{proof}

\begin{theorem}\label{app:theorem_conv}
The Magnus expansion $\Omega(t)$, defined by $\mathcal{T}e^{-i\alpha\int_0^tH_1(s)ds}=e^{\Omega(t)}$ and 
the series in
\cref{eq:Magnus_map}, converges for $|\alpha|\int_0^t\|H_1(s)\|ds\leq 1.08687\dots$.
\end{theorem}\label{theo:convergence_magnus}

\begin{proof}
 Starting from the definition of the Magnus operator \cref{eq:Magnus_map}, the triangle inequality gives
\begin{align}
    \|\Omega(t)\|\leq |\alpha|\int_0^t\sum_{k=0}^\infty\frac{|b_k|}{k!}(2\|\Omega(s)\|)^k\|H_1(s)\|ds=|\alpha|\int_0^tg(2\|\Omega(s)\|)\|H_1(s)\|ds.
\end{align}
As $g$ is a nondecreasing positive function in the interval $[0,2\pi)$, we can apply \cref{lemma_bihari}, giving
\begin{align}
    \|\Omega(t)\|\leq \frac{1}{2}G^{-1}\left(2|\alpha|\int_0^t\|H_1(s)\|ds\right).
\end{align}
This implies that $\|\Omega(t)\|$ is bounded as long as
\begin{align}
|\alpha|\int_0^t\|H_1(s)\|ds\leq \frac{1}{2}G(2\pi)=\frac{1}{2}\int_0^{2\pi}\frac{ds}{2+\frac{x}{2}(1-\cot(x/2))}=1.08687\dots,
\end{align}
as claimed.
\end{proof}

The proof of the Magnus-\THR approximation theorem (\cref{thm:magnus_thrift}) uses \cref{lem:bound_omega}, which we now prove.

\begin{lemma}\label{lem:bound_omega_app}
For $1\leq l$,
$\|\tilde{\Omega}_l(t)\|\leq \frac{1}{2}x_l(2\int_0^t\|H_1(s)\| ds)^l$, where $x_l$ is the coefficient of $s^l$ in the expansion of $G^{-1}(s)=\sum_{m=1}^\infty x_m s^m$, the inverse function of $G(s)=\int_0^s(2+\frac{x}{2}(1-\cot(x/2))^{-1}dx$.
\end{lemma}

\begin{proof} 
We proceed by induction. First, as $\tilde{\Omega}_1(t)= -i\int_0^t H_1(s) ds$, we have
\begin{align}
\| \tilde{\Omega}_1(t)\|\leq \int_0^t\|H_1(s)\| ds=\frac{x_1}{2}\left(2\int_0^t\|H_1(s)\| ds\right)
\end{align}
with $x_1=1$. The induction hypothesis is $\|\tilde{\Omega}_l(t)\|\leq \frac{1}{2}x_l(2\int_0^t\|H_1(s)\| ds)^l$ for  $1\leq l\leq n$. 
To prove the induction step, we integrate \cref{eq:tilde_omega} and use the triangle inequality, leading to
\begin{align}
 \label{eq:extension}
\|\tilde{\Omega}_{n+1}(t)\| &\leq \sum_{j=1}^{n}\frac{|b_{j}|}{j!}\sum_{\substack{k_{1}+k_{2}+\dots+k_{j}=n\\k_1,k_2,\dots,k_j\geq 1}}\int_{0}^{t}2^{j}\prod_{m=1}^{j}\|\tilde{\Omega}_{k_{m}}(s)\|\|H_1(s)\|ds,\\\nonumber
&=\sum_{j=1}^{n}\frac{|b_{j}|}{j!}\int_{0}^{t}2^{j}\hat{B}_{n,j}(\|\tilde{\Omega}_{1}(s)\|,\dots,\|\tilde{\Omega}_{n-j+1}(s)\|)\|H_{1}(s)\|ds,
\end{align}
where we have introduced the ordinary Bell polynomials \cite{Bell1934,comtet1974advanced}, defined by
\begin{align}\label{eq:bell_polys}
\hat{B}_{n,j}(x_1,x_2,\dots,x_{n-j+1}):=\frac{1}{n!}\left.\frac{\partial^{n}}{\partial\alpha^{n}}\left(\sum_{k=1}^\infty\alpha^kx_k\right)^j\right|_{\alpha=0}=\sum_{\substack{k_{1}+k_{2}+\dots+k_{j}=n\\k_1,k_2,\dots,k_j\geq 1}}\prod_{m=1}^jx_{k_m}.
\end{align}
Using the induction hypothesis on \cref{eq:extension} and that $\hat{B}_{n,j}(rx_1,r^2x_2\dots,r^{n-k+1}x_{n-k+1})=r^n\hat{B}_{n,j}(x_1,x_2\dots,x_{n-k+1})$, which follows from the definition \cref{eq:bell_polys}, we have
\begin{align}
\|\tilde{\Omega}_{n+1}(t)\|&\leq \left(\int_{0}^{t}\left(2\int_0^s\|H_1(y)\| dy\right)^n\|H_1(s)\|ds\right)\sum_{j=1}^{n}\frac{|b_{j}|}{j!}\hat{B}(x_1,\dots,x_{n-j+1}),\\
&=\left(\frac{1}{2}\int_{0}^{t}\frac{d}{ds}\frac{(2\int_{0}^{s}\|H(x)\|dx)^{n+1}}{n+1}ds\right)\sum_{j=1}^{n}\frac{|b_{j}|}{j!}\hat{B}_{n,j}(x_{1},\dots,x_{n-j+1})\quad\mbox{using the chain rule,}\\\label{eq:induction_step}
&=\frac{(2\int_{0}^{t}\|H(x)\|dx)^{n+1}}{n+1}\frac{1}{2}\sum_{j=1}^{n}\frac{|b_{j}|}{j!}\hat{B}_{n,j}(x_{1},\dots,x_{n-j+1})\quad\mbox{using the fundamental theorem of calculus}.
\end{align}
To finish the proof, we show that the factor $\frac{1}{(n+1)}\sum_{j=1}^{n}\frac{|b_{j}|}{j!}\hat{B}_{n,j}(x_{1},\dots,x_{n-j+1})$ corresponds to the coefficient of $z^{n+1}$ in the series expansion $G^{-1}(z)=\sum_{m=1}^\infty z^mx_m$,
given that $\{x_j\}_{j=1}^n$ are also coefficients of $G^{-1}$. That can be shown as follows:
\begin{align}\label{eq:Bell_bell}
X_{n+1}:=\frac{1}{n+1}\sum_{j=1}^{n}\frac{|b_{j}|}{j!}\hat{B}_{n,j}(x_{1},\dots,x_{n-j+1})=\frac{1}{(n+1)!}\sum_{j=1}^{n}|b_{j}|B_{n,j}(1!x_{1},2!x_{2},\dots,(n-j+1)!x_{n-k+1})
\end{align}
where we used the relation
\begin{align}
    \frac{n!}{j!}\hat{B}_{n,j}(x_{1},\dots,x_{n-j+1})=B_{n,j}(1!x_{1},2!x_{2},\dots,(n-j+1)!x_{n-k+1})
\end{align}
between the ordinary Bell polynomials $\hat{B}_{n,k}$ and the exponential Bell polynomials $B_{n,k}$ \cite{comtet1974advanced}. Now note that $G^{-1}(z)=\sum_{n=1}^\infty z^nx_n$ implies $\frac{d^{n}G^{-1}(0)}{dz^{n}}=n!x_{n}$, so we can write \cref{eq:Bell_bell} as
\begin{align}\nonumber
X_{n+1}&=\frac{1}{(n+1)!}\sum_{j=1}^{n}|b_{j}|B_{n,j}\left(\frac{dG^{-1}(0)}{dz},\frac{d^2G^{-1}(0)}{dz^2},\dots,\frac{d^{n-j+1}G^{-1}(0)}{dz^{n-j+1}}\right),\\\label{eq:X_n}
&=\frac{1}{(n+1)!}\sum_{j=1}^{n}\frac{d^jg(0)}{dz^j}B_{n,j}\left(\frac{dG^{-1}(0)}{dz},\frac{d^2G^{-1}(0)}{dz^2},\dots,\frac{d^{n-j+1}G^{-1}(0)}{dz^{n-j+1}}\right),
\end{align}
with $g(z)=\sum_{j=0}^\infty \frac{|b_j|}{j!}z^j=2+\frac{z}{2}(1-\cot(z/2))$. Finally, using the derivative rule for inverse functions $\frac{dG^{-1}(z)}{dz}=\frac{1}{G'(G^{-1}(z))}$ and the definition of $G(z)=\int_0^z(g(s))^{-1}ds$, we have $\frac{dG^{-1}(z)}{dz}=g(G^{-1}(z))$. In general,
\begin{align}\label{eq:Faa}
    \frac{d^{n+1}}{dz^{n+1}}(G^{-1}(z))=\frac{d^{n}}{dz^{n}}(g(G^{-1}(z)))=\sum_{k=1}^{n}\frac{d^j}{dz^j}(g(G^{-1}(z))B_{n,k}\left(\frac{dG^{-1}(z)}{dz},\frac{d^{2}G^{-1}(z)}{dz^{2}},\dots,\frac{d^{n-k+1}G^{-1}(z)}{dz^{n-k+1}}\right)
\end{align}
where we have used Fa\`{a} di Bruno's identity for the generalised chain rule \cite{Craik2005}. Comparing \cref{eq:X_n} and \cref{eq:Faa}, we find
\begin{align}
X_{n+1}=\frac{1}{(n+1)!}\frac{d^{n+1}}{dz^{n+1}}(G^{-1}(0)),
\end{align}
which is by definition $x_{n+1}$. Going back to \cref{eq:induction_step}, this implies
\begin{align}
\|\tilde{\Omega}_{n+1}(t)\|\leq \frac{x_{n+1}}{2}\left(2\int_{0}^{t}\|H(x)\|dx\right)^{n+1}=\frac{1}{2(n+1)!}\frac{d^{n+1}}{dz^{n+1}}(G^{-1}(0))\left(2\int_{0}^{t}\|H(x)\|dx\right)^{n+1}.
\end{align}
This proves the induction step and hence the lemma.
\end{proof}

\section{Circuit details for numerical implementations}
\label{app:sec:circuit_depths_numerical_implementations}

In this section we discuss the circuit depth for the both Trotter and \THR algorithms, using arbitrary 2-qubit gates, for the transverse-field Ising model (1D and 2D cases), 1D Heisenberg model, and 1D Fermi-Hubbard model. We consider a Hamiltonian of the form $H = H_0 + \alpha H_1$, where $H_0$ is a sum of single-qubit terms (unless otherwise specified); $H_1 = \sum_{j=1}^{K}h_j$, with each $h_j$ containing terms acting on disjoint qubits; and $\alpha \ll 1$. Exponentials of the terms in $H_1$, $e^{-ih_jt}$, can therefore be implemented simultaneously with $\mathcal{N}_j$ arbitrary 2-qubit gates. For all the models we consider, we have $\mathcal{N}_j = \mathcal{N}$, independent of $j$. 

\subsection{General facts about product formulas}

\subsubsection{Trotter formulas}

The first-order Trotter approximation (Trotter 1) for the time-evolution operator $U = e^{-iHt}$ is
\begin{equation}
    \label{app:eq:trotter1}
    \mathcal{S}_1(t) = P_1^K(t),
\end{equation}
with 
\begin{equation}
    P_a^b(z) = \left(\prod_{j=a}^{b-1} e^{-ih_jz}\right) e^{-iH_0 z} e^{-ih_bz}. 
\end{equation}
Since $e^{-iH_0t}$ only requires single-qubit gates,  \cref{app:eq:trotter1} can be implemented with $K\mathcal{N}$ layers of arbitrary 2-qubit gates. 

The second-order Trotter approximation (Trotter 2) can be written as
\begin{equation}
    \label{app:eq:trotter2}
    \mathcal{S}_2(t) = P_1^K(t/2)P_K^1(t/2) = P_1^{K-1}(t/2) e^{-ih_Kt} P_{K-1}^1(t/2),
\end{equation}
and can be implemented with $(2K-1)\mathcal{N}$ layers of arbitrary 2-qubit gates. Note that if the number of Trotter layers is $N>1$, one can merge the last exponential of the $(i-1)$st step with the first of the $(i)$th step, giving a total arbitrary 2-qubit gate depth of $[(2K-2)N + 1]\mathcal{N}$.

The fourth-order Trotter approximation (Trotter 4) can be obtained from \cref{app:eq:trotter2} as \cite{Hatano_2005} 
\begin{equation}
    \label{app:eq:trotter4}
    \mathcal{S}_4(t) = \mathcal{S}_2(s_2t)^2 \mathcal{S}_2((1-4s_2)t) \mathcal{S}_2(s_2t)^2, \qquad \text{with } s_2 := (4-\sqrt[3]{4})^{-1}.
\end{equation}
The final term of each $\mathcal{S}_2(z)$ can be merged with the first term of the following $\mathcal{S}_2(z)$, so \cref{app:eq:trotter4} can be implemented with $[5(2K-2)+1]\mathcal{N} = (10K - 9)\mathcal{N}$ layers of arbitrary 2-qubit gates. As in the Trotter 2 case, if the number of Trotter layers is $N>1$, one can merge the last time-evolution operator of the $(i-1)$st step with the first of the $(i)$th step. This gives a total arbitrary 2-qubit gate depth of $[5(2K-2)N + 1]\mathcal{N}$.

Finally, the optimised eighth-order Trotter approximation (optimised Trotter 8) is given by Eq.~(15) in \cite{morales2022greatly}
\begin{equation}
    \label{app:eq:trotter8}
    \mathcal{S}_8(t) = \left(\prod_{j=1}^{m}\mathcal{S}_2(\omega_{m-j+1}t)\right)\mathcal{S}_2(\omega_{0}t)\left(\prod_{j=1}^{m}\mathcal{S}_2(\omega_{j}t)\right),
\end{equation}
with $m=7$. Similarly to the previous case, one obtains that \cref{app:eq:trotter8} can be implemented with $[15(2K-2)+1]\mathcal{N} = (30K - 29)\mathcal{N}$ layers of arbitrary 2-qubit gates. If the number of Trotter layers is $N>1$, one can merge the last time-evolution operator of the $(i-1)$st step with the first of the $(i)$th step. This gives a total arbitrary 2-qubit gate depth of $[15(2K-2)N + 1]\mathcal{N}$.

\subsubsection{The ``small A'' formula of Omelyan et al.}
\label{app:subsubsec:omelyan}
In Ref.~\cite{Omelyan_2002_optimized}, \citeauthor{Omelyan_2002_optimized} derive an optimised fourth-order product formula for a Hamiltonian $H = H_0 + \alpha H_1$ with $\alpha \ll 1$. The error achieved by this formula scales as $O(\alpha^2 t^5) + O(\alpha t^7)$. This implies that there exists a regime for small time $t$ in which this formula achieves a scaling in $\alpha$ similar to THRIFT. 

For a Hamiltonian $H = H_0 + \alpha H_1$, \citeauthor{Omelyan_2002_optimized}'s optmised formula can be written as~\cite{Omelyan_2002_optimized, Ostmeyer_2023_trotter}
\begin{equation}
    \label{app:eq:omelyan}
    U_O(t) = e^{-ia_1H_0t} e^{-ib_1\alpha H_1t} e^{-ia_2H_0t} e^{-ib_2\alpha H_1t} e^{-ia_3H_0t} e^{-ib_2\alpha H_1t} e^{-ia_2H_0t} e^{-ib_1\alpha H_1t} e^{-ia_1H_0t},
\end{equation}
where the numerically determined coefficients are 
\begin{align}
a_1 &= 0.5316386245813512,\nonumber\\
b_1 &= -0.04375142191737413,\nonumber\\
a_2 &= -0.3086019704406066,\\
b_2 &= \frac{1}{2}-b_1,\nonumber\\
a_3 &= 1-2\sum_{i=1}^{2}a_i.\nonumber
\end{align}

Since, for all the systems considered in this work, we have $H = \sum_{k=1}^{\Lambda} h_k$ with $\Lambda\geq2$, we use the generalisation of \cref{app:eq:omelyan} to Hamiltonians with an arbitrary number of terms given in Eq.~(3) of Ref.~\cite{Ostmeyer_2023_trotter}. In particular, 
\begin{equation}
    \label{app:eq:omelyan_gen}
    U_O(t) = \left(\prod_{k=1}^{\Lambda}e^{-ic_1h_kt}\right)\left(\prod_{k=\Lambda}^{1}e^{-id_1h_kt}\right)...\left(\prod_{k=1}^{\Lambda}e^{-ic_4h_kt}\right)\left(\prod_{k=\Lambda}^{1}e^{-id_4h_kt}\right),
\end{equation}
where $c_i = a_i - d_{i-1}$ (with $d_0 = 0$) and $d_i = b_i - c_{i}$.

\subsubsection{\THR formulas}

The circuit depths for implementing a \THR approximation of order $p$ are the same as the corresponding Trotter approximation applied to the original Hamiltonian with the rearrangement
\begin{align}
H = \sum_{j=1}^{K}[(H_0+h_j)- H_0] = \sum_{j=1}^{K}[h'_j-H_0],
\end{align}
where we assume that the exponential of each $h'_j := H_0 + h_j$ can be implemented with $\mathcal{N}'_j = \mathcal{N}'$ arbitrary 2-qubit gates. Note that particular care is required in cases where $H_0$ contains terms acting on more that one qubit, as in the 1D Fermi-Hubbard model case discussed in \cref{app:subsubsec:1D_FermiHubbard}. 

\subsubsection{Magnus-\THR formulas}

The first-order Magnus-\THR formula is given by \cref{eq:approx_Magnus} with $k = 1$,
\begin{equation}
    \mathcal{S}^{\mathrm{Magnus}}_1(t) = e^{-iH_0t} e^{\Omega^{[1]}(t)},
\end{equation}
with
\begin{equation}
    \Omega^{[1]}(t) = -i\int_{0}^{t}H_1(t_1)dt_1,
\end{equation}
where $H_1(t) = e^{iH_0t} H_1 e^{-H_0t}$. In general, we can write $H_1(t) = \sum_{j=1}^{P} f_j(t)\tilde{h}_{j}$, where the exponential of each $\tilde{h}_j$ can be implemented with $\tilde{\mathcal{N}}_j$ layers of arbitrary 2-qubit gates. Hence
\begin{equation}
    \int_{0}^{t}H_1(t_1) dt_1= \sum_{j=1}^{P}  \left(\int_{0}^{t}f_j(t_1) dt_1 \right) \tilde{h}_{j} = \sum_{j=1}^{P} F_j(t) \tilde{h}_{j} = H^{\mathrm{Magnus}}_{1}(t)
\end{equation}
and
\begin{equation}
    \label{app:eq:magnus1}
    \mathcal{S}^{\mathrm{Magnus}}_1(t) = e^{-iH_0t} e^{-iH^{\mathrm{Magnus}}_{1}(t)}.
\end{equation}
Approximating the last term by a first-order Trotter formula, \cref{app:eq:magnus1} can be implemented with $\sum_{j=1}^{P}\tilde{\mathcal{N}}_j$ layers of arbitrary 2-qubit gates.

The second-order Magnus-\THR formula is given by \cref{eq:approx_Magnus} with $k = 2$,
\begin{equation}
    \label{app:eq:Magnus2}
    \mathcal{S}^{\mathrm{Magnus}}_2(t) = e^{-iH_0t} e^{\Omega^{[2]}(t)},
\end{equation}
with
\begin{equation}\label{app:eq:Omega2}
    \Omega^{[2]}(t) = -i\int_{0}^{t}H_1(t_1)dt_1 - \frac{1}{2}\int_{0}^{t}dt_1\int_{0}^{t_1}dt_2 [H_1(t_1), H_1(t_2)].
\end{equation}

In this case, we can write $[H_1(t_1), H_1(t_2)] = \sum_{i,j=1}^{P} f_{i}(t_1)f_{j}(t_2) \tilde{h}_i \tilde{h}_j$, and therefore Eq.~\eqref{app:eq:Omega2} becomes
\begin{equation}
    \Omega^{[2]}(t) = -i\sum_{j=1}^{P'} g_j(t) \bar{h}_j,
\end{equation}
with $P'\leq P^2$. The time-evolution operator of each term $\bar{h}_j$ can be implemented with arbitrary 2-qubit gate depth $\bar{\mathcal{N}}_j$. Approximating $e^{\Omega^{[2]}(t)}$ by a second-order Trotter formula and assuming for simplicity that $\bar{\mathcal{N}}_j = \bar{\mathcal{N}}$ for all $j$, we find that a single Trotter layer of \cref{{app:eq:Magnus2}} can be implemented with arbitrary 2-qubit gate depth $(2P'-1)\bar{\mathcal{N}}$ and $N>1$ Trotter layers with arbitrary 2-qubit gate depth $[(2P'-2)N+1]\bar{\mathcal{N}}$. Note that, in general, the terms $\bar{h}_j$ may contain multi-qubit terms and therefore $\bar{\mathcal{N}}$ depends on the specific model. In the next section we discuss the case of the 1D \TFIM. 

\subsection{Applications to specific models}
\subsubsection{1D transverse-field Ising model}
\label{app:subsubsec:1D_TFIM}
For the 1D transverse-field Ising model, we have $H_{\mathrm{TFIM}} = H_0 + J H_1$ with $H_0$ acting on single qubits only and $H_1 = h^{\mathrm{even}}_{XX} + h^{\mathrm{odd}}_{XX}$ (hence $K=P=2$), with $h^{\mathrm{even/odd}}_{XX} = \sum_{j\ \mathrm{even/odd}} X_{j}X_{j+1}$. The time-evolution operators associated with such terms can be implemented with $\mathcal{N} = 1$ layer of arbitrary 2-qubit gates (or two layers of CNOT gates) each. Since the cost to implement the time-evolution operator of $h^{\mathrm{even/odd}}_{XX} + H_0$ is the same as $h^{\mathrm{even/odd}}_{XX}$ in terms of both arbitrary 2-qubit gates and CNOT gates, the circuit depth for a $p$th-order \THR formula is the same as for the corresponding Trotter formula. 

\begin{figure}
    \centering
    \includegraphics[width=0.7\linewidth]{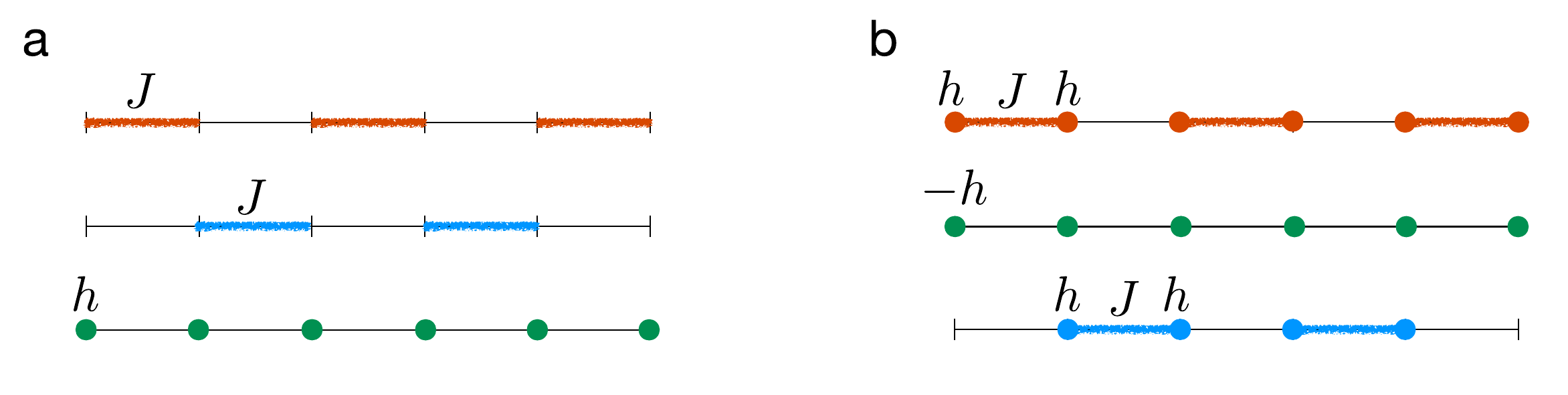}
    \caption{Partitions for implementing the first-order (a) Trotter and (b) \THR formulas in the 1D TFIM and 1D Heisenberg models. (a) One layer of the first-order Trotter approximation for these models is obtained by approximating $e^{-iHt}$ as $e^{-ih^{\mathrm{odd}}_{1}} e^{-ih^{\mathrm{even}}_{1}t} e^{-iH_0t} $. (b) One layer of the first-order \THR approximation is obtained by approximating $e^{-iHt}$ as $e^{-i(h^{\mathrm{odd}}_{1} + H_0)t}  e^{iH_0t} e^{-i(h^{\mathrm{even}}_{1}+H_0)t} e^{-iH_0t} $. Here, $h^{\mathrm{even/odd}}_{1} = h^{\mathrm{even/odd}}_{XX}$ for the 1D transverse-field Ising model and $h^{\mathrm{even/odd}}_{1} = h^{\mathrm{even/odd}}_{XX+YY+ZZ}$ for the 1D Heisenberg model. In each row of both panels, all the single-qubit (dots) and 2-qubit (thick lines) terms are implemented simultaneously.}
    \label{app:fig:1d_tfim_partitions}
\end{figure}

For \citeauthor{Omelyan_2002_optimized}'s optimised small $A$ formula, we have $\Lambda=3$, and we identify $h_1 = H_0$, $h_2 = \alpha h_{XX}^{\mathrm{even}}$, and $h_3 = \alpha h_{XX}^{\mathrm{odd}}$. Recalling that $H_0$ can be implemented via single-qubit gates and merging the last exponential inside a bracket in \cref{app:eq:omelyan_gen} with the first exponential of the following one, we find that the arbitrary 2-qubit gate cost to implement $N$ layers of \cref{app:eq:omelyan_gen} is $12N$ (corresponding to $24N$ layers of CNOT gates).

\begin{figure}
    \centering
    \includegraphics[width=\linewidth]{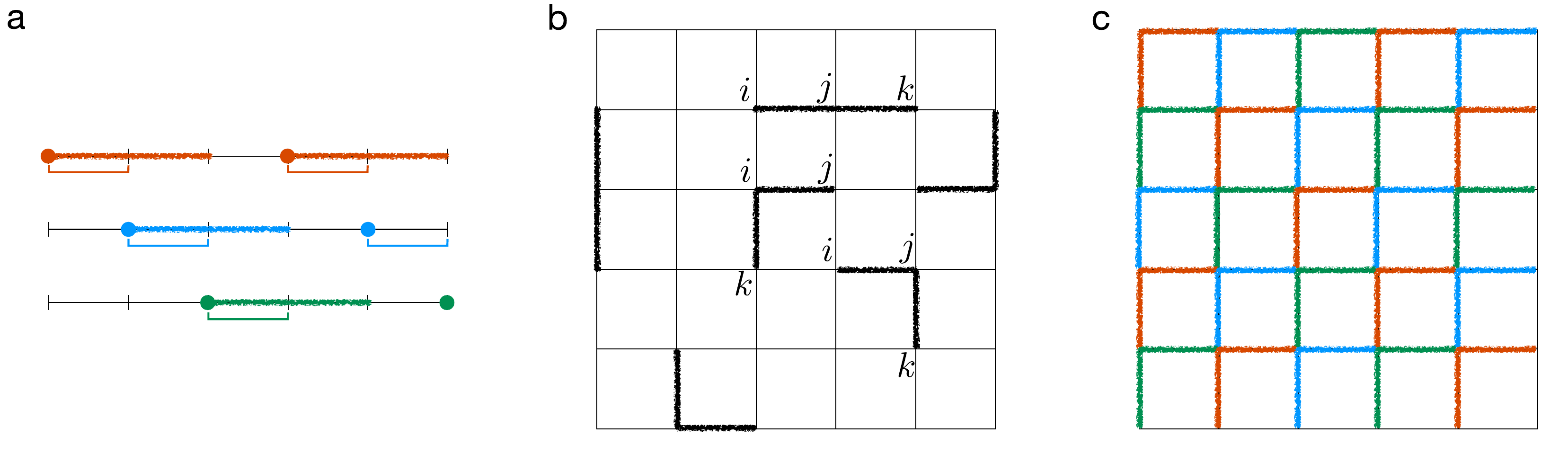}
    \caption{
    Partitions for implementing the second-order Magnus-\THR formula in the 1D (a) and 2D (b,c) \TFIM. (a) One of the possible groupings of the terms in $\Omega^{[2]}(t)$ in \cref{app:eq:Omega2_1DTFIM}. In each one of the three subgroups, all the single-qubit (dots), 2-qubit (thin lines), and three-qubit (thick lines) terms are implemented simultaneously. (b) In the 2D \TFIM, there are six possible configurations for each of the three-qubit terms in $\Omega^{[2]}(t)$ in \cref{app:eq:Omega2_1DTFIM}. (c) In order to implement $e^{\Omega^{[2]}(t)}$, each of the six possible configurations can be split into three layers of terms that can be implemented simultaneously. 
    }
  \label{app:fig:2d_tfim_partitions}
\end{figure}

For Magnus-\THR 1 we have 
\begin{equation}
    \label{app:eq:H1_magnus_1DTFIM}
    H_1(t) = f_{XX}(t) H_{XX} + f_{YY}(t) H_{YY} + f_{XY+YX}(t) H_{XY+YX},
\end{equation}
with $H_{XX} = \sum_{j} X_{j} X_{j+1}$, $H_{YY} = \sum_{j} Y_{j} Y_{j+1}$, $H_{XY+YX} = \sum_{j} \left(X_{j} Y_{j+1} + Y_{j} X_{j+1}\right)$, and $f_i(t)$ time-dependent coefficients. Similarly to the previous cases, $H_1(t)$ can be split into even/odd contributions, each of which can be implemented with one layer of arbitrary 2-qubit gates (or two CNOT gates). Hence, we have $P=2$ and $\tilde{\mathcal{N}}_j = \tilde{\mathcal{N}} = 1$: $N>1$ layers of the Magnus-\THR 1 formula in \cref{app:eq:magnus1} can be implemented with $2N$ layers of arbitrary 2-qubit gates (or $4N$ layers of CNOT gates).
For Magnus-\THR 2, \cref{app:eq:Omega2} can be written as
\begin{align}
    \label{app:eq:Omega2_1DTFIM}
    \Omega^{[2]}(t) &\propto f_{XX}(t)H_{XX} + f_{YY}(t)H_{YY} + f_{XY+YX}(t)H_{XY+YX} \nonumber\\
    &\quad+ f_{XZY+YZX}(t)H_{XZY+YZX} + f_{XZX}(t)H_{XZX} + f_{YZY}(t)H_{YZY}\nonumber\\
    &\quad+ \text{single-qubit terms}.
\end{align}
Here, $H_{XZY+YZX} = \sum_{j} X_{j}Z_{j+1}Y_{j+2} + Y_{j}Z_{j+1}X_{j+2}$, $H_{XZX} = \sum_{j} X_{j}Z_{j+1}X_{j+2}$, $H_{YZY} = \sum_{j} Y_{j}Z_{j+1}Y_{j+2}$, and the various $f_i(t)$ denote the corresponding time-dependent coefficients. Since the terms in the second line of \cref{app:eq:Omega2_1DTFIM} act on three qubits, $\Omega^{[2]}(t)$ has to be split into three groups as shown in \cref{app:fig:2d_tfim_partitions}(a). Moreover, one can show numerically that the time-evolution operator of each group can be implemented with 3 layers of arbitrary 2-qubit gates (corresponding to 9 layers of CNOT gates). Hence, the second-order Magnus-\THR formula for the 1D transverse-field Ising model one has $P' = 3$ and $\bar{\mathcal{N}}_j = \bar{\mathcal{N}} = 3$, corresponding to an arbitrary 2-qubit gate depth of $12N + 3$ (and CNOT gate depth of $36N + 9$).

\subsubsection{2D transverse-field Ising model}
\label{app:subsubsec:2D_TFIM}
For the 2D \TFIM we have $H_{\mathrm{TFIM}} = H_0 + J H_1$, with $H_{1} = H^{\mathrm{h}}_{XX} + H^{\mathrm{v}}_{XX} = h^{\mathrm{h, even}}_{XX} + h^{\mathrm{h, odd}}_{XX} + h^{\mathrm{v,even}}_{XX} + h^{\mathrm{v,odd}}_{XX}$, with the h and v superscripts denoting horizontal and vertical terms, respectively. Hence, we have $K=4$. The time-evolution operator corresponding to each term can be implemented with a layer of $\mathcal{N} = 1$ arbitrary 2-qubit gates (or 2 CNOT gates). Similarly to the 1D case, one can implement the time-evolution operators of $h^{\mathrm{h/v,even/odd}}_{XX} + H_0 $ occurring in \THR formulas with the same cost. 

For \citeauthor{Omelyan_2002_optimized}'s optimised small $A$ formula we have $\Lambda=5$, with $h_1 = H_0$, $h_2 = \alpha h_{XX}^{\mathrm{h,even}}$, $h_3 =\alpha  h_{XX}^{\mathrm{h,odd}}$, $h_4 = \alpha h_{XX}^{\mathrm{v,even}}$, and $h_5 = \alpha h_{XX}^{\mathrm{v,odd}}$ (see \cref{app:subsubsec:2D_TFIM}). With similar arguments as in the 1D case, we find that the arbitrary 2-qubit gate cost to implement $N$ layers of \cref{app:eq:omelyan_gen} is $28N$ (or $56N$ layers of CNOT gates).

In Magnus-\THR 1, $H_1(t)$ has the same form as \cref{app:eq:H1_magnus_1DTFIM} and can be split into four terms as the original $H_1$. Therefore, we find $P=4$ and $\tilde{\mathcal{N}}_j = \tilde{\mathcal{N}} = 1$: $N>1$ layers of the Magnus-\THR 1 formula in \cref{app:eq:magnus1} can be implemented with $4N$ layers of arbitrary 2-qubit gates (or $8N$ layers of CNOT gates). 

The implementation of the second-order Magnus-\THR approximation requires more care. The functional form of $\Omega^{[2]}(t)$ is the same as in \cref{app:eq:Omega2_1DTFIM}, but each of the three-qubit Hamiltonians $H_{XZY+YZX}$, $H_{XZX}$, and $H_{YZY}$ has now a 2D nature. For instance, $H_{XZX} = \sum_{\langle i, j\rangle} \sum_{k\in\mathrm{neigh}(\{i,j\})} X_{i}Z_{j}X_{k}$: here, $i,j$ are nearest-neighbors and $k$ is a nearest-neighbor of either $i$ or $j$. Hence, for a given choice of $i,j$, there are 2 linear (vertical and horizontal) and 4 two-dimensional ``L''-shaped independent configurations (see \cref{app:fig:2d_tfim_partitions}(b)). The time-evolution operators corresponding to each of these terms can be implemented in 3 layers as shown in \cref{app:fig:2d_tfim_partitions}(c). Then, $P' = 18$. In turn, we numercally verified that each layer can be implemented with $\bar{\mathcal{N}} = 3$ arbitrary-two qubit gates (or 9 CNOT gates). The overall arbitrary 2-qubit (CNOT) gate depth to implement $N>1$ steps is therefore $102N +3 $ ($306N + 9$). 

\subsubsection{1D Heisenberg model}
\label{app:subsubsec:1D_Heisenberg}

Similarly to the 1D transverse-field Ising model, for the 1D Heisenberg model we have $H_{\mathrm{Heisenberg}} = H_0 + J H_1$ with $H_0$ acting on single qubits only and $H_1 = H_{XX} + H_{YY} + H_{ZZ} = h^{\mathrm{even}}_{XX+YY+ZZ} + h^{\mathrm{odd}}_{XX+YY+ZZ}$ (hence $K=2$), with $h^{\mathrm{even/odd}}_{XX+YY+ZZ} = \sum_{j\ \mathrm{even/odd}} \left(X_{j}X_{j+1} + Y_{j}Y_{j+1} + Z_{j}Z_{j+1}\right)$. Therefore, the circuit depths for Trotter and \THR formulas for the 1D Heisenberg model can be obtained by following the same steps as the 1D \TFIM discussed in \cref{app:subsubsec:1D_TFIM}. The same holds for \citeauthor{Omelyan_2002_optimized}'s optimised small $A$ formula, where $\Lambda=3$ and we identify $h_1 = H_0$, $h_2 = \alpha h_{XX+YY+ZZ}^{\mathrm{even}}$, and $h_3 = \alpha h_{XX+YY+ZZ}^{\mathrm{odd}}$.
In particular, the arbitrary 2-qubit gate depths for the various formulas are the same, while to obtain the CNOT gate depths, one has to take into account that the time-evolution operator associated with $ h^{\mathrm{even/odd}}_{XX+YY+ZZ} $ uses 3 layers of CNOT gates.

\subsubsection{1D Fermi-Hubbard model}
\label{app:subsubsec:1D_FermiHubbard}

The Hamiltonian of the Fermi-Hubbard model can be written as  $H_{\mathrm{FH}} = H_0 + \hop H_1 $, with $H_0 = H_{\mathrm{int}}$ and $H_1 = H_{\mathrm{hop}} = h^{\mathrm{even}}_{\mathrm{hop}} + h^{\mathrm{odd}}_{\mathrm{hop}}$, with $h^{\mathrm{even/odd}}_{\mathrm{hop}} = -\sum_{\sigma}\sum_{i\ \mathrm{even/odd}}\left(c^{\dagger}_{i,\sigma} c_{i+1,\sigma} + c^{\dagger}_{i+1,\sigma} c_{i,\sigma}\right)$. The time-evolution operator corresponding to each term of this decomposition can be implemented with one layer of arbitrary 2-qubit gates (and 2 layers of CNOT gates). Hence, we find $K=3$ and $\mathcal{N}=1$. 

For \citeauthor{Omelyan_2002_optimized}'s optimised small $A$ formula we have $\Lambda=3$ with $h_1 = H_{\mathrm{int}}$, $h_2 = \alpha h_{\mathrm{hop}}^{\mathrm{even}}$, and $h_3 = \alpha h_{\mathrm{hop}}^{\mathrm{odd}}$ (see \cref{app:subsubsec:1D_FermiHubbard}). In contrast to the previous cases, here $H_0$ can be implemented with one layer of arbitrary 2-qubit gates. Hence, the overall arbitrary 2-qubit gate cost to implement $N$ layers of \cref{app:eq:omelyan_gen} is $16N + 1$ (corresponding to $32N+2$ layers of CNOT gates). Note that, to obtain this number, one needs to merge the last term $e^{-id_4H_{\mathrm{int}}t}$ of the $i$th Trotter layer with the first term $e^{-ic_1H_{\mathrm{int}}t}$ of the $(i+1)$st layer.

\begin{figure}
  \includegraphics[width=\linewidth]{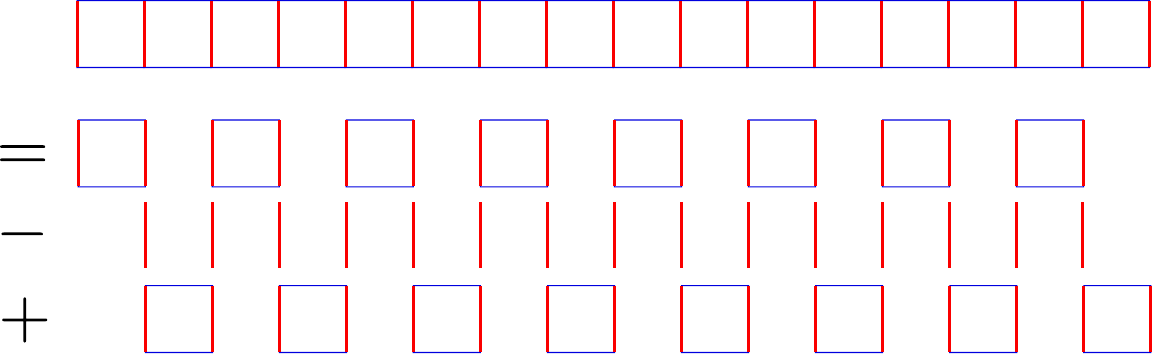}
  \caption{Partition of terms for \THR in the 1D Fermi-Hubbard model. Vertical lines (red) correspond to the on-site interaction, while horizontal lines (blue) correspond to the hopping terms. Taking $H_0=\sum_jn_{j\uparrow} n_{j\downarrow}$ leads to a partition where 4 qubit gates are needed. }\label{fig:FH_THR}
\end{figure}

Finally, there are some additional considerations to obtain the 2-qubit gate depth for \THR formulas. In this case, $H_0$ is not a single-qubit Hamiltonian and implementing $e^{\pm i H_0 t}$ requires $\mathcal{N}'_0 = 1$ layer of arbitrary 2-qubit gates (corresponding to 2 layers of CNOT gates). As shown in \cref{fig:FH_THR}, each of the $K=2$ \THR partitions $[H_0 + H^{\mathrm{even/odd}}]$ consist of terms acting on four qubits and implementing $e^{-i (H_0 + H^{\mathrm{even/odd}}) t}$ requires 3 layers of arbitrary 2-qubit gates (and 6 layers of CNOT gates). Hence, $\mathcal{N}'_1 = 3$. The overall arbitrary 2-qubit (CNOT) gate depth can be computed by taking into account both these facts. For instance, $N>1$ layers of the second-order THRIFT formula can be implemented with arbitrary 2-qubit gate depth $[(2K-2)\mathcal{N}'_1 + 2\mathcal{N}'_0)]N + \mathcal{N}'_1 = 8N + 3$.

\section{Additional numerical results} \label{sec:additional_numerics}

In \cref{sec:numerical_results} we showed the 2-qubit gate depth $d$ to achieve 
a fixed precision $\epsilon$ for different system sizes $L$ at evolution time $T = L$.
In this section we provide a more detailed analysis by showing 
the results at different values of the small parameter $\alpha$, performing 
weighted linear regression to the power laws describing the depth $d$ as a 
function of $L$, and comparing the power laws thus obtained to the theoretically 
expected results.

For later reference we note that for ordinary $k$th-order Trotter methods,
the depth to achieve error $\epsilon$ scales for evolution time 
$T$ in a system of size $L$ scales as
\begin{equation}
  d_{\textnormal{trotter},k} = O \left(\epsilon^{-\frac{1}{k}} \alpha^{\frac{1}{k}} L^{\frac{1}{k}} T^{1 + \frac{1}{k}} \right)
  \label{eq:trotter_depth_requirement}
\end{equation}
if we choose a splitting of the Hamiltonian that has $H_0$ as one term and all 
other terms scale linear with $\alpha$.
For a $k$th-order \THR formula, almost the same is true; the only difference 
is that the commutator bounds now give a factor of $\alpha^2$, so the depth scales as 
\begin{equation}
  d_{\textnormal{thrift},k} = O \left(\epsilon^{-\frac{1}{k}} \alpha^{\frac{2}{k}} L^{\frac{1}{k}} T^{1 + \frac{1}{k}} \right).
  \label{eq:thrift_depth_requirement}
\end{equation}
These two expressions follow simply from combining the ordinary 
Trotter error bounds, or \THR error bounds given in \cref{thm:thrift}, with 
the analysis from \cref{sec:commutator_scaling}.
Because we only consider geometrically local Hamiltonians,
\cref{eq:trotter_depth_requirement,eq:thrift_depth_requirement} hold 
with $\epsilon$ denoting the worst-case error, as in \cref{fig:tfim_thrift_depth_requirements_worstcase},
as well as when $\epsilon$ is the average case error, as in
\cref{fig:heisenberg_thrift_depth_requirements_averagecase,fig:fh_thrift_depth_requirements_t=0.0625_averagecase}, by the same analysis done for Theorem 2 in \cite{zhao2022hamiltonian}.

\subsection{Transverse-field Ising model} \label{sec:additional_numerics_tfim}

\begin{figure}
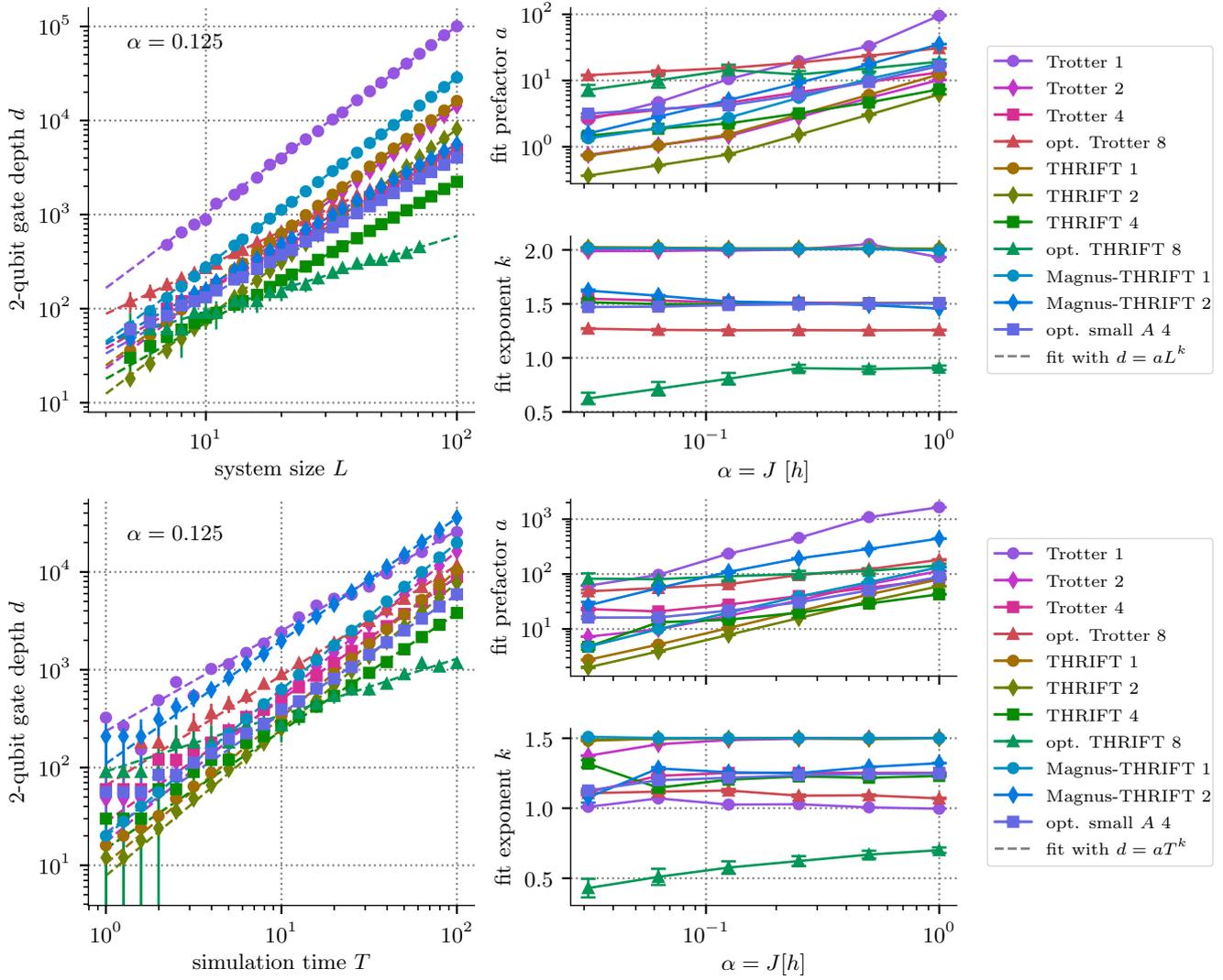

  \begin{center}
    \input{figures/tfim_tds_algo_fits_1d.pgf}
    \input{figures/tfim_tds_algo_fits_2d.pgf}
  \end{center}
  \caption{Performance scaling for the 1D (top) and 2D (bottom) transverse field Ising model. (left) The same data as in
    \cref{fig:tfim_thrift_depth_requirements_worstcase} to
    provide context to the fit parameters shown on the right. (right) Fit
    parameters of a power law $d = a L^k$ or $d = a T^k$, respectively, to the
    data shown on the left for different values of $\alpha$, obtained via
    weighted linear regression.
    }
  \label{fig:tfim_tds_algo_fits}
\end{figure}

In \cref{fig:tfim_tds_algo_fits} we analyse in more detail the 2-qubit depth required to 
achive a worst-case error $\norm{U - U_{\mathrm{exact}}} < 0.01$ for the different 
TDS methods as a function of system size $L$ and evolution time $T$ (1D case)
or only evolution time $T$ (2D case). On the left
we show the 2-qubit gate depth $d$ at fixed $\alpha = 1/8$ and see that it 
is well described by a power law of the form $d = a L^k$ (1D) or $d = a T^k$ (2D).
On the right we plot the prefactor $a$ and exponent $k$ as a function of $\alpha$.
In the 1D case, the exponents of the second- and fourth-order methods match the theoretically 
expected values of 2 and 1.5 very well. The same is true for the 
optimised eighth-order formula where the exponent is $\approx 1.125$ for all 
$\alpha$. The exponents of the first- and second-order methods match. This is because the \TFIM Hamiltonian 
and $H_1(t)$ both can be decomposed into only two terms that are exactly implementable,
in which case the first-order Trotter formula has the same scaling as the second-order 
formula. The fit exponent of the optimised \THR 8 formula, on the other hand, 
does not match the theoretically expected value and is below $1$ for all $\alpha$,
despite the very accurate fits shown on the left. Instead we find that $a$ scales roughly as 
$\alpha^{\frac{2}{k}}$ for second-, fourth-, and eighth-order 
\THR \emph{and} Trotter methods, although the prefactors $a$ of the \THR methods 
are always below those of the corresponding Trotter method. 
Again, the first-order methods behave similarly to the second-order methods and 
$a$ is roughly linear in $a$ for both Magnus-\THR methods.

In the 2D case the fit exponents do not fall as nicely into distinct groups,
but we observe again that, with the notable exception of Trotter 1, all 
first- and second-order methods have exponent $k \approx 1.5$ as theoretically 
expected for second-order methods. The fourth-order methods have $k \approx 1.25$,
again in line with theoretical expectations. Trotter 1 and the optimised \THR 8
formula, on the other hand, deviate substantially 
from the theoretical expectation with $k \approx 1$ and $k \approx 0.5$, respectively. This 
suggests that the optimised \THR 8 formula can be used to fast forward the 
\TFIM. While the 1D \TFIM is integrable,
this is more surprising in the 2D case and may be an artifact of the fairly small
system size considered here. For the prefactors $a$, we find the same as in 1D: 
they have $d$ scaling like $\alpha^{\frac{2}{k}}$ for \THR \emph{and} Trotter methods,
i.e., as theoretically expected for the \THR methods.

\subsection{1D Heisenberg model} \label{sec:additional_numerics_heisenberg}

\begin{figure}
  \begin{center}
    \input{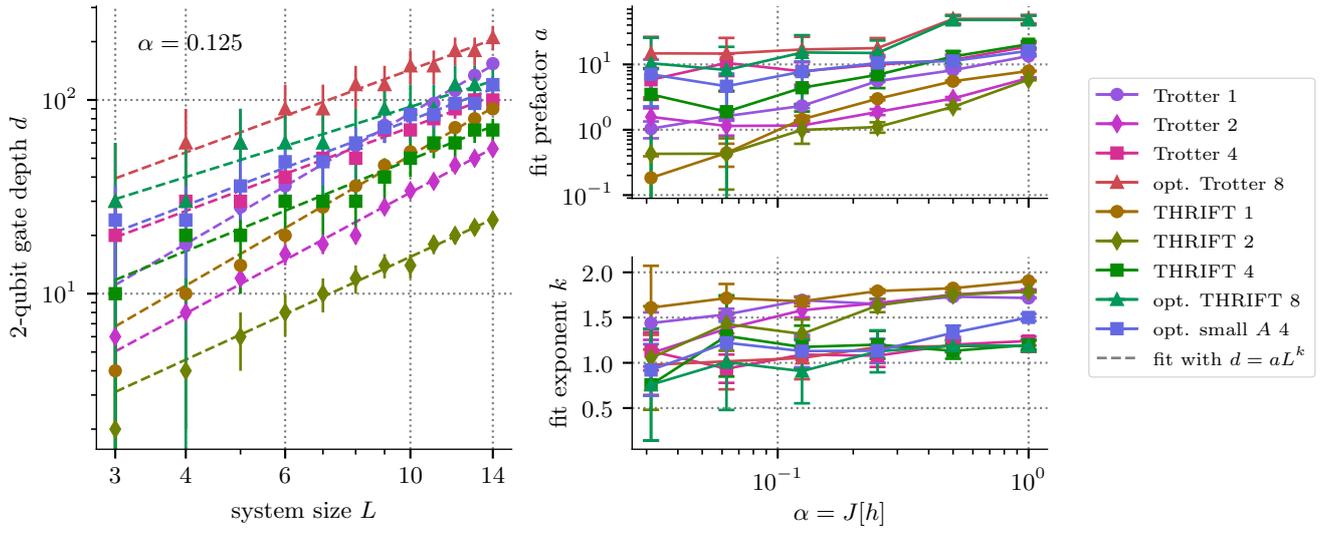}
  \end{center}
  \caption{Performance scaling for the 1D     Heisenberg chain. (left) The same data as in \cref{fig:heisenberg_thrift_depth_requirements_averagecase} to provide context for the fit parameters shown on the right. (right) Fit parameters for the data shown on the left for different values of $\alpha$, obtained via weighted linear regression.}
  \label{fig:heisenberg_tds_algo_fits_1d}
\end{figure}

In \cref{fig:heisenberg_tds_algo_fits_1d} we analyse the 2-qubit gate depth to achieve an average infidelity
$\mathbb{E}_{\{\ket{x}\}}[1 - |\braket{x|U_{\mathrm{exact}}^\dagger U|x}|^2] \leq  0.01$
as a function of the system size $L$, evolution time $T$, and 
interaction strength $J = \alpha$. On the left we show the 2-qubit gate
depth at fixed $\alpha = \frac{1}{8}$, which is well described by a power law
of the form $d = a L^k$. We find that this remains true for different choices of $\alpha$,
where the coefficients $a$ and $k$ depend on $\alpha$. On the right we
show the coefficients obtained via weighted linear regression as a function
of $\alpha$. While the situation is not as clear cut as for the \TFIM in 
\cref{fig:tfim_tds_algo_fits}, the algorithms (with maybe the exception of 
the optimised small $A$ method) still appear to fall into 
two groups: the first- and second-order methods, for which (at least for 
larger $\alpha \gtrsim 0.2$) $k \approx 1.75$, and the higher-order methods, for which 
(again, at least for $\alpha \gtrsim 0.2$) $k \approx 1.25$.

\subsection{1D \FH model} \label{sec:additional_numerics_fh}

\begin{figure}
  \begin{center}
    \input{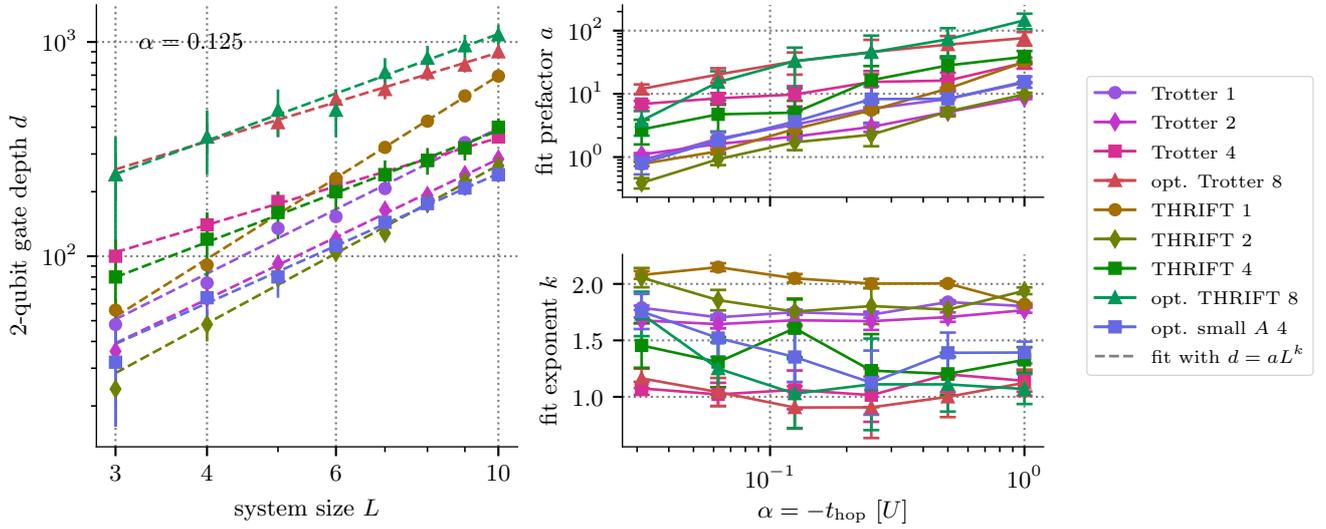}
  \end{center}
  \caption{Performance scaling for the 1D     Fermi-Hubbard model. (left) The
    same data as in
    \cref{fig:fh_thrift_depth_requirements_t=0.0625_averagecase} to provide
    context for the fit parameters shown on the right. (right) Fit parameters
    to the data shown on the left for different values of $\alpha$. The error
    bars are the fit uncertainties when taking the error bars from the left as
    the uncertainties of the original data.
    }
  \label{fig:fh_tds_algo_fits_1d}
\end{figure}

The same analysis done for the \TFIM and Heisenberg model in 
\cref{fig:tfim_tds_algo_fits,fig:heisenberg_tds_algo_fits_1d} is repeated for 
the \FH model in \cref{fig:fh_tds_algo_fits_1d}. Again, we use the
average infidelity $\mathbb{E}_{\{\ket{x}\}}[1 - |\braket{x|U_{\mathrm{exact}}^\dagger U|x}|^2] \leq  0.01$ as a figure of merit to be able to reach larger system sizes in our
simulations. 
Again, we find robust power laws for the 2-qubit depth to get the 
average infidelity below threshold as we increase the system size $L$ and scale 
the evolution time as $T = L$, as exemplified for $\alpha = 1/16$ in 
the left of \cref{fig:fh_tds_algo_fits_1d}. On the right we plot the exponents $k$ 
and prefactors $a$ of that power law as a function of $\alpha$. 

As in the case of Heisenberg model (\cref{fig:heisenberg_tds_algo_fits_1d}),
the algorithms do not fall as neatly into groups with different exponents 
as for the \TFIM (\cref{fig:tfim_tds_algo_fits}).
Trotter 1 and 2 have $k \approx 1.75$ for all $\alpha$, and Trotter 4 and the optimised 
Trotter 8 formula have $1 \lesssim k \lesssim 1.25$ for all $\alpha$, but also 
fairly large uncertainties. For \THR 1 and 2, $k$ varies between $2.25$ and $1.75$,
and for \THR 4 and 8, it decreases with $\alpha$ from $k \approx 1.75$ at $\alpha = 1/32$
to $k \approx 1.25$ at $\alpha = 1$.

\subsection{TFIM and Heisenberg model with strong interactions \texorpdfstring{$\alpha = 1$}{alpha=1}} \label{sec:additional_numerics_large_alpha}

\Cref{fig:tfim_tds_algo_comparison,fig:heisenberg_tds_algo_comparison} indicate 
that the \THR methods perform well for the \TFIM and Heisenberg model not only
in the theoretically expected $\alpha \ll 1$ limit, but also for $\alpha \sim 1$.
In \cref{fig:tfim_thrift_depth_requirements_worstcase_large_alpha,fig:heisenberg_thrift_depth_requirements_averagecase_large_alpha}
we show that this is indeed the case by repeating the numerics done for $\alpha = 1/8$ in 
\cref{fig:tfim_thrift_depth_requirements_worstcase,fig:heisenberg_thrift_depth_requirements_averagecase}, now taking the larger value
$\alpha = 1$. We find that for the \TFIM, the \THR circuits use lower 
depth than Trotter circuits to achieve a desired precision even at $J = h$, and 
that for the Heisenberg model, the depths are very similar for the 
\THR and Trotter methods.

\begin{figure}
\centering
\begin{minipage}[t]{.49\textwidth}
  \centering
  \input{figures/tfim_thrift_depth_requirements_J=1.0_worstcase.pgf}
  \captionof{figure}{
    2-qubit gate depth to achieve $\norm{U - U_{\mathrm{exact}}} \leq 0.01$ for
    the different TDS algorithms for a field strength of $J=1$ and
    evolution time $T = L$, for a $1 \times L$ Ising chain with transverse field
    $h=1$. In contrast to \cref{fig:tfim_thrift_depth_requirements_worstcase},
    we have $\alpha = 1$, so \cref{thm:thrift} 
    does not predict that \THR methods should outperform Trotter methods. Nevertheless, 
    \THR uses shallower circuits to achieve the desired precision than
    the corresponding Trotter methods.
  }
  \label{fig:tfim_thrift_depth_requirements_worstcase_large_alpha}
\end{minipage}%
\hfill
\begin{minipage}[t]{.49\textwidth}
  \centering
  \input{figures/heisenberg_thrift_depth_requirements_t=1.0_averagecase.pgf}
  \captionof{figure}{
    2-qubit depth to achieve average infidelity 
    $\mathbb{E}_{\{\ket{x}\}}[1 - |\braket{x|U_{\mathrm{exact}}^\dagger U|x}|^2] \leq 0.01$
    for the different TDS algorithms for
    a $1 \times L$ Heisenberg chain with field strength of $J = 1$ and evolution time $T = L$.
    In contrast to \cref{fig:heisenberg_thrift_depth_requirements_averagecase},
    we have $\alpha = 1$ here, so \cref{thm:thrift} does not 
    predict that \THR methods should outperform Trotter methods. Nevertheless, \THR 
    uses almost the same circuit depth as the corresponding 
    Trotter methods to achieve the target precision.
  }
  \label{fig:heisenberg_thrift_depth_requirements_averagecase_large_alpha}
\end{minipage}
\end{figure}

\clearpage
\printbibliography

\end{document}